\documentclass[12pt, draft]{article}
\usepackage{amsmath, amssymb}
\usepackage[active]{srcltx}
\usepackage{rangecite}

\makeatletter
\renewcommand\section{\@startsection {section}{1}{\z@}%
                                   {-3.5ex \@plus -1ex \@minus -.2ex}%nn
                                   {2.3ex \@plus.2ex}%
                                   {\normalfont\large\bfseries}}
\renewcommand\subsection{\@startsection{subsection}{2}{\z@}%
                                     {-3.25ex\@plus -1ex \@minus -.2ex}%
                                     {1.5ex \@plus .2ex}%
                                     {\normalfont\bfseries}}
\makeatother

\newcommand{\nc}{\newcommand}
 \numberwithin{equation}{section}

 \nc{\ra}{\rightarrow}

 \def\dA{\,\delta{\!A_D}\,}
\def\dAt{\,\delta{\!A_b}\,}

\def\rep#1{{{\bf{#1}}}}

 \nc{\sigb}{\ol{\sigma}}
\nc{\thetab}{\ol{\theta}}
\def\id{\mathbb{I}}
\def\GLD{G_{\rm LD}}
\def\Dloop{{\Delta_{\rm 1-loop}\,}}
\nc{\cDslash}{\ensuremath \raisebox{0.025cm}{\,\slash}\hspace{-0.30cm} {\cD\,}}
\nc{\cDslashb}{\ensuremath \raisebox{0.025cm}{\,\slash}\hspace{-0.30cm} {\bar{\cD}}\,}

 \def\dpsi{\delta{\psi}}
\def\dpsib{\delta{\bar\psi}}
\def\dPsi{\delta{\Psi}}
\def\dPsib{\delta{\bar\Psi}}

\def\al{\alpha}
\def\a{\alpha}
\def\b{\beta}

\def\eps{\epsilon}
\def\e{\epsilon}
\nc{\ve}{\varepsilon}
\def\gam{\gamma}

\def\om{\omega}
\nc{\vphi}{\varphi}
\def\tha{\theta}

\def\sig{\sigma}

\def\Gam{\Gamma}

\def\Om{\Omega}
\def\Sig{\Sigma}

\newcommand{\beq}{\begin{equation}}
\newcommand{\eeq}{\end{equation}}
\newcommand{\beqnn}{\begin{equation*}}
\newcommand{\eeqnn}{\end{equation*}}
\newcommand{\bean}{\begin{eqnarray*}}
\newcommand{\eean}{\end{eqnarray*}}

\nc{\bea}{\begin{eqnarray}}
\nc{\eea}{\end{eqnarray}}
\nc{\be}{\begin{equation}}
\nc{\ee}{\end{equation}}
\nc{\cA}{{\cal A}}
\nc{\cB}{ \cal B}

\def\cD{{\cal D}}

\nc{\cF}{{\cal F}}
\nc{\cG}{{\cal G}}
\nc{\cg}{{\cal g}}

\nc{\cL}{{\cal L}}
\nc{\M}{{\cal M}}
\nc{\cM}{{\cal M}}
\def\N{{\cal N}}

\def\cO{{\cal O}}

\nc{\cQ}{{\cal Q}}
\nc{\cR}{{\cal R}}

\def\T{{\cal T}}

\def\cZ{{\cal Z}}
\nc{\BB}{{\mathbb B}}
\nc{\CC}{{\mathbb C}}
\nc{\DD}{{\mathbb D}}
\nc{\EE}{{\mathbb E}}
\nc{\FF}{{\mathbb F}}
\nc{\GG}{{\mathbb G}}
\nc{\HH}{{\mathbb H}}
\nc{\JJ}{{\mathbb J}}
\nc{\RR}{{\mathbb R}}
\nc{\MM}{{\mathbb M}}
\nc{\PP}{{\mathbb P}}
\nc{\QQ}{{\mathbb Q}}
\nc{\ZZ}{{\mathbb Z}}
\nc{\CP}{{\CC\PP}}
\nc{\calone}{{\mathbb 1}}

\newcommand{\IC}{\mathbb{C}}

\newcommand{\IP}{\mathbb{P}}

\newcommand{\IR}{\mathbb{R}}

\newcommand{\IZ}{\mathbb{Z}}

\nc{\half}{\frac{1}{2}}
\nc{\qrt}{\frac{1}{4}}
\nc{\del}{\partial}

\nc{\delbar}{\bar\partial}
\nc{\Spin}{\operatorname{Spin}}
\nc{\SO}{\operatorname{SO}}
\nc{\OO}{{\cal{O}}}
\nc{\Sp}{{\rm Sp}}

\nc{\com}[2]{{ \left[ #1, #2 \right] }}
\nc{\acom}[2]{{ \left\{ #1, #2 \right\} }}
\nc{\rr}{\rightarrow}
\nc{\p}{\partial}
\nc{\LT}{{\LL_\T}}
\nc{\Tr}{{\rm Tr}}
\nc{\tr}{{\rm tr}}
\def\com#1#2{{ \left[ #1, #2 \right] }}
\def\acom#1#2{{ \left\{ #1, #2 \right\} }}
\nc{\wt}{\widetilde}
\nc{\tKT}{\widetilde{K3}}
\nc{\ttha}{\tilde{\theta}}
\nc{\tphi}{\tilde{\phi}}
\nc{\tPhi}{\tilde{\Phi}}
\nc{\tpsi}{\tilde{\psi}}
\nc{\tgam}{\tilde{\gam}}
\nc{\tGam}{\tilde{\Gam}}
\nc{\tSig}{\tilde{\Sig}}
\nc{\tc}{\tilde c}

\nc{\te}{\tilde e}
\nc{\tg}{\tilde g}
\nc{\tj}{\tilde j}
\nc{\tp}{\widetilde{p}}
\nc{\tq}{\widetilde{q}}
\nc{\ts}{{\tilde s}}
\nc{\tz}{\tilde z}
\nc{\tA}{{\tilde A}}
\nc{\tD}{{\tilde D}}
\nc{\tE}{{\tilde E}}
\nc{\tG}{{\tilde G}}
\nc{\tH}{{\tilde H}}
\nc{\tM}{{\tilde M}}
\nc{\tN}{{\tilde N}}
\nc{\tP}{{\tilde P}}
\nc{\tQ}{{\tilde Q}}
\nc{\tS}{\tilde{S}}
\nc{\tF}{\tilde{{\cal F}}}
\nc{\tX}{\widetilde{X}}
\nc{\hb}{\hat b}
\nc{\hc}{\hat c}
\nc{\hd}{\hat d}
\nc{\he}{\hat e}
\nc{\hf}{\hat f}
\nc{\hg}{\hat g}
\nc{\hh}{\hat h}
\nc{\hp}{\hat p}
\nc{\hw}{\hat w}
\nc{\hx}{\hat x}
\nc{\hy}{\hat y}
\nc{\hz}{\hat z}
\nc{\hA}{\widehat{A}}
\nc{\hE}{\widehat{E}}
\nc{\hH}{\widehat{H}}
\nc{\hF}{\widehat{F}}
\nc{\hJ}{\widehat{J}}
\nc{\tK}{\widetilde{K}}
\nc{\hM}{\widehat M}

\nc{\ha}{\widehat \alpha}
\nc{\hphi}{\hat{\phi}}
\nc{\hpsi}{\hat{\psi}}
\nc{\hgam}{\hat{\gam}}
\nc{\hPhi}{\hat{\Phi}}
\nc{\hPsi}{\hat{\Psi}}
\nc{\hGam}{\hat{\Gam}}

\nc{\w}{\wedge}

\nc{\ol}{\overline}
\nc{\abar}{{\ol{a}}}
\nc{\bbar}{{\ol{b}}}
\nc{\cbar}{{\ol{c}}}
\nc{\ebar}{{\ol{e}}}
\nc{\ibar}{{\ol{\imath}}}
\nc{\jbar}{{\ol{\jmath}}}
\nc{\kbar}{{\ol{k}}}
\nc{\lbar}{{\ol{l}}}
\nc{\mbar}{{\ol{m}}}
\nc{\nbar}{{\ol{n}}}
\nc{\ubar}{{\ol{u}}}
\nc{\vbar}{{\ol{v}}}
\nc{\wbar}{{\ol{w}}}
\nc{\xbar}{{\ol{x}}}
\nc{\ybar}{{\ol{y}}}
\nc{\zbar}{{\ol{z}}}

\nc{\Eb}{\ol{E}}
\nc{\Qb}{\ol{Q}}
\nc{\Pb}{\ol{P}}
\nc{\Jb}{\ol{J}}
\nc{\Wb}{\ol{W}}
\nc{\Xb}{{\overline X}}
\nc{\Yb}{{\overline Y}}
\nc{\Zb}{{\overline Z}}

\nc{\epsbar}{\ol{\epsilon}}
\nc{\lambar}{\ol{\lambda}}
\nc{\psibar}{\ol{\psi}}
\nc{\Psibar}{\ol{\Psi}}
\nc{\psib}{\ol{\psi}}
\nc{\Psib}{\ol{\Psi}}

\nc{\phibar}{\ol{\phi}}

\nc{\Phibar}{\ol{\Phi}}
\nc{\chibar}{\ol{\chi}}
\nc{\ombar}{\ol{\om}}
\nc{\Ombar}{\ol{\Om}}

\nc{\epsb}{\ol{\epsilon}}
\nc{\lamb}{\ol{\lambda}}
\nc{\gammab}{\ol{\gamma}}
\nc{\Gammab}{\ol{\Gamma}}
\nc{\lambdab}{\ol{\lambda}}
\nc{\phib}{\ol{\phi}}
\nc{\Phib}{\ol{\Phi}}
\nc{\chib}{\ol{\chi}}
\nc{\omb}{\ol{\om}}
\nc{\Omb}{\ol{\Om}}
\nc{\alphabar}{\ol{\alpha}}
\nc{\betabar}{\ol{\beta}}
\nc{\etab}{\ol{\eta}}

\nc{\bah}{{\mathbf {\hat{A}}}}
\nc{\bX}{{\mathbf X}}

\nc{\dal}{\dot{\al}}
\nc{\thab}{\bar{\theta}}
\nc{\thal}{\theta^{\al}}
\nc{\thdal}{\bar{\theta}^{\dal}}

\nc{\thsigthm}{\tha \sigma^m \thab}
\nc{\thsigthn}{\tha \sigma^n \thab}

\nc{\Dal}{D_{\al}}
\nc{\Ddal}{\bar{D}_{\dal}}
\nc{\CDal}{{\cal D}_{\al}}
\nc{\CDdal}{\bar{\cal D}_{\dal}}

\nc{\eq}[1]{(\ref{#1})}
\nc{\non}{\nonumber}
\nc{\comment}[1]{\noindent{\bf #1}}
\nc{\fcomment}[1]{\footnote{{\bf #1}}}
\nc{\bwcomment}[1]{\footnote{{\bf[\textcolor{red}{ #1}}]}}
\nc{\xs}{\not\!\!X}
\nc{\ps}{\not\!\!P}
\nc{\dif}{{d}}
\nc{\equ}{{\rm eq}}

\nc{\AdS}{{\rm AdS}}
\nc{\vol}{{\rm vol}}
\nc{\Ainf}{A_{\infty}}
\nc{\End}{{\rm End}}
\nc{\Ext}{{\rm Ext}}
\nc{\Hom}{{\rm Hom}}
\nc{\IIB}{{\rm IIB}}
\nc{\Pic}{{\rm Pic}}
\nc{\SL}{{\rm SL}}
\nc{\GL}{{\rm GL}}
\nc{\Ker}{{\rm Ker}}
\nc{\diag}{{\rm diag}}
\nc{\FD}{{F_{D}}}
\nc{\FA}{{F_{A}}}
\nc{\bra}[1]{\langle{#1}|}
\nc{\ket}[1]{|{#1}\rangle}
\nc{\braket}[2]{\langle{#1}|{#2}\rangle}
\nc{\sect}[1]{Section~\ref{#1}}
\nc{\fig}[1]{Fig.~\ref{#1}}
\nc{\chap}[1]{Chapter~\ref{#1}}
\nc{\Dslash}{\ensuremath \raisebox{0.025cm}{\slash}\hspace{-0.32cm} D}

\nc{\no}{\!:\!\!}
\nc{\bpm}{\begin{pmatrix}}
\nc{\epm}{\end{pmatrix}}
 \nc{\bi}{\begin{itemize}}
 \nc{\ei}{\end{itemize}}
 \nc{\ben}{\begin{enumerate}}
 \nc{\een}{\end{enumerate}}

\newcommand{\C}[1]{$(\ref{#1})$}

\def\psib{\ol{\psi}}

\def\Z{\mathbb{Z}}

\def\SO{\operatorname{SO}}

\def\SU{\operatorname{SU}}
\def\U{\operatorname{U}}

\nc{\rank}{{\rm rank}}
\nc{\pr}{{\rm pr}}

\nc{\tom}{\tilde{\om}}
\nc{\tOm}{\tilde{\Om}}

\def\M{\mathcal{M}}
\def\inv{^{-1}}
\def\e{\epsilon}
\def\bz{{\ol{z}}}

\def\bZ{\ol{Z}}
\def\Zb{\ol{Z}}
\def\bw{\ol{w}}

\newcommand{\vev}[1]{\langle {#1} \rangle}

\def\dZ{\delta{Z}}

\def\dZb{\,\delta{\bar{Z}}}

\def\Zb{\ol{Z}}
\def\bw{\bar{w}}

\begin{document}
\begin{titlepage}

\begin{center}

%{September 26, 2005}
\today \hfill         \phantom{xxx}% \hfill EFI-xx-xx\\
 				    \hfill DAMTP-2011-109

\vskip 2 cm {\Large \bf Monopole--Instantons in M2-brane Theories}\non\\
\vskip 1.25 cm { Emil Martinec$^{a}$\footnote{ejmartin@uchicago.edu} and Jock McOrist$^{b}$\footnote{j.mcorist@damtp.cam.ac.uk}}\non\\
{\vskip 0.5cm $^{a}$ Enrico Fermi Institute,
University of Chicago, Chicago, IL 60637, USA\non\\ \vskip 0.2cm
$^{b}$ Department of Applied Mathematics and
Theoretical Physics, Centre for Mathematical Sciences, Wilberforce
Road, Cambridge, CB3 0WA, UK\non\\ }

\end{center}
\vskip 2 cm

\begin{abstract}
\baselineskip=18pt
We study monopole-instantons in M2-brane theories, focussing on the ABJM class of Chern-Simons gauge theories coupled to matter. We calculate calculate explicitly the 8-fermion term in the effective action induced by these monopole-instantons, and discuss their role in resolving a classical singularity in the moduli space. The results are compared with monopole-instantons in N=8 3d SYM and D-brane theories, as well the dual supergravity description as a membrane scattering process.

\end{abstract}

\end{titlepage}

\pagestyle{plain}
%\baselineskip=18pt
% Try a wider skip
\baselineskip=19pt
%\newpage
%\tableofcontents
%\newpage

%%%%%%%%%%%%%%%%%%%%%%%%%%%%%%%%%%%%%%%%%%%%%%%%%%%%%%%%%%%%%%%%%%%%%%%%%%%%%%

\section{Introduction}
The  work of Bagger-Lambert \cite{Bagger:2006sk,Bagger:2007jr,Bagger:2007vi}, Gustavsson \cite{Gustavsson:2008dy,Gustavsson:2007vu} and ABJM \cite{Aharony:2008ug} represents an important step forward in our understanding of the conformal field theory describing coincident M2-branes. While the work of BLG provided the initial breakthrough in understanding the conformal field theory describing multiple M2-branes, its description is seemingly limited to two M2-branes in a certain orbifold background. The ABJM theory improved on this, proposing to describe $N$ M2-branes probing $\IC^4/\IZ_k$. Both theories enjoy many common qualitative features and ingredients. Most importantly, they give explicit Lagrangian descriptions of the conformal field theory, and hence open up the possibility of explicitly computing quantities peculiar to M2-brane theories. An example is understanding the dynamics and scattering of M2-branes, much in the way \cite{Douglas:1996yp} explored scattering of D-branes. To make progress in this directions it is important to understand the quantum corrected moduli space of the ABJM and BLG theories.

In this note we will explore two related aspects of the ABJM moduli space. The first is the appearance of a distinguished locus in the classical moduli space. When any two M2-branes lie along this locus, we find new massless off-diagonal states, even though the M2-branes may be separated arbitrarily far apart. Furthermore, there are no enhanced gauge symmetries that would typically be associated with such a singular locus. This is in contrast to D-brane theories, where the only time one finds singularities in the moduli space and associated massless states is when a pair of D-branes coincide with an associated gauge symmetry enhancement. Physically, the massless D-brane states and symmetry enhancement are ascribed to open strings becoming light, and a question arises: what is the physical interpretation of the anomalously light M2-brane states?

These off-diagonal states appear to have been largely overlooked in the literature. Some exceptions include \cite{Lambert:2008et} who speculated the analogous massless excitations in BLG theories describe a type of three-prong object, which might be related to the $N^{3/2}$ entropy scaling of M2-brane SCFTs. In the context of ABJM, \cite{Berenstein:2008dc} labelled them `membrane bits', proposing that a pair of M2-branes are connected by a single membrane bit. A simple scaling argument suggested that a membrane bit has two spatial dimensions and wraps the M-theory circle so that when two M2-branes are separated along the M-theory circle, the membrane bits become massless. However, both papers largely ignore the role of quantum corrections in the dynamics of these excitations.

In the low-energy effective action on the Higgs branch, supersymmetry dictates that the first quantum correction appears at the four-derivative level, or equivalently an  8-fermion coupling \cite{Dorey:1997tr,Polchinski:1997pz,Paban:1998mp}. It is generated by monopole--instantons,  the dimensional reduction of monopoles to three Euclidean dimensions. Monopole--instantons in ABJM have been discussed in \cite{Hosomichi:2008ip}, where a finite-energy BPS solution to the equations of motion was constructed. However, an explicit calculation of the influence of monopole-instantons on the moduli space dynamics is lacking. Our goal here is to both rectify this as well as discuss how monopole--instantons affect the distinguished locus and the corresponding massless off-diagonal modes discussed above. Although we will focus on the ABJM theory, similar conclusions to apply for the BLG theory, as well as generalisations of ABJM to M2-branes probing non-compact toric Calabi-Yau's.

The outline for the remainder of this paper is the following. In the next section we will review ABJM and its classical moduli space. We will identify the singular locus in ABJM, in its generalisations, and in the BLG theory. In section 3, we will review some generalities of monopole--instantons and how they appear in ABJM. In section 4 we will discuss their role in resolving the singular locus and discuss open questions.  Three appendices discuss the generalization to related theories, the one-loop fluctuation determinant, and the zero mode analysis.

%%%%%%%%%%%%%%%%%%%%%%%%%%%%%%
%%%%%%%%%%%%%%%%%%%%%%%%%%%%%%

\section{ABJM on the Higgs Branch}
The $\N=6$ ABJM theory is a superconformal Chern-Simons matter theory defined on a three-manifold $\Sigma$ with a $\U(N)\times \U(N)$ gauge group  coupled to bifundamental matter. The gauge fields are denoted by $A_{(1)}$, $A_{(2)}$ and have Chern-Simons levels $(k,-k)$. The bifundamental matter fields are composed of four complex scalars $Z^P$ and their fermionic partners $\psi_P$. Both fields transform in the $(N,\overline{N})$ representation of the gauge group. There is a global $\SU(4)$ R-symmetry under which the scalars $Z^P$ and fermions $\psi_P$ transform in the $4$. Further details of our notation are given in the appendix. The Lagrangian of \cite{Aharony:2008ug} is given by
\begin{eqnarray}
S &=& S_{KE} + S_{\rm int} + S_V + S_{CS},\label{action1}
\end{eqnarray}
where the individual components of the action are given by
\begin{align}
S_{KE} =& - \int_\Sigma d^3x\, \tr(D_\mu Z^{P} D^\mu \Zb_P) + i\tr(\psib^{P}\gamma^\mu D_\mu \psi_P),\cr
 S_{V} =& - \int_\Sigma d^3x\, V(Z), \cr
 S_{\rm int} =& \frac{2\pi i }{k} \int_\Sigma d^3x\, \Big\{\tr(2\psib^{P} \psi_Q \Zb_P Z^Q - 2\psib^{P} Z^Q \Zb_P \psi_Q - \psib^{P}\psi_P \Zb_Q Z^Q + \psib^{P} Z^Q \Zb_Q \psi_P)+\cr
&- \varepsilon_{PQRS} \tr(\psib^{P} Z^Q \psib^{R} Z^S) +\varepsilon^{PQRS} \tr(\bar Z_P \psi_Q \bar Z_R \psi_S)\Big\},\cr
S_{CS} =& \frac{k}{4\pi} \int_\Sigma  \omega_{CS}(A_{(1)}) - \omega_{CS}( A_{(2)}).
\label{csaction}
\end{align}
We have written the action in Lorentz signature, though will eventually switch to Euclidean signature for the instanton calculation. The covariant derivative for the scalars is given by
\be
D Z^P = dZ^P - i A_{(1)} Z^P + i Z^P A_{(2)},\label{covderiv}
\ee
while the Chern-Simons form is given by
\be
\omega_{CS}(A) = \tr \Big( A \w dA - \frac{2i}{3} A\w A \w A\Big).
\ee
The $U(N)\times U(N)$ gauge transformations act as
\be
\begin{split}
&Z \rightarrow LZM^{-1}, \quad \Zb \rightarrow L\inv \Zb M, \cr
&A_{(1)} \rightarrow L A_{(1)} L\inv - i dL  L\inv, \quad A_{(2)} \rightarrow M A_{(2)} M\inv - i dM M\inv,
\end{split}
\label{eqn:gauge1}
\ee
where $L,M$ are $U(N)$ matrices. The Chern-Simons form transforms as
\be
\begin{split}
&\omega_{CS}(A_{(1)})\rightarrow \omega_{CS}(A_{(1)}) -i d \Big[\tr\big(A_{(1)} L\inv dL\big)\Big] - \frac{1}{3} \tr (L\inv dL)^3,\cr
&\omega_{CS}(A_{(2)})\rightarrow \omega_{CS}(A_{(2)}) -i d \Big[\tr\big(A_{(2)}  M\inv dM\big)\Big] - \frac{1}{3} \tr (M\inv dM)^3,\cr
\end{split}
\label{cstransformation}
\ee
The bosonic potential $V(Z)$ can be written has a sum of squares
\be
V = \frac{2\pi^2}{3k^2} \tr(\Upsilon_{R}^{PQ}\overline{\Upsilon}^{R}_{PQ}),\label{pot}
\ee
where
\be
\Upsilon^{PQ}_R = (2 Z^P \Zb_R Z^Q - \delta^Q_R Z^P \Zb_S Z^S - \delta^P_R Z^S \Zb_S Z^Q) - (P \leftrightarrow Q).\label{ups}
\ee
The supersymmetry transformations are given by
\begin{align}
&\delta Z^P = - i \eta^{PQ} \psi_Q,\cr
&\delta \psi_P = \Big[ \gamma^\mu D_\mu Z^R - \frac{4\pi}{3k} (Z^{[Q}\Zb_{Q} Z^{R]})\Big] \eta_{RP} + \frac{8\pi}{3k} (Z^Q \Zb_P Z^R)\eta^{RQ} - \frac{4\pi}{3k} \e_{PQRS} (Z^Q \Zb_E Z^R) \eta^{DE},\cr
&\delta A_{(1)\,\mu} = \frac{2\pi i}{k} \big( \eta_{PQ} \gamma_\mu Z^P \bar\psi^Q + \eta^{PQ} \gamma_\mu \psi_Q \Zb_P\big), \cr
& \delta A_{(2)\,\mu} = \frac{2\pi i}{k} \big( \eta_{PQ} \gamma_\mu  \bar\psi^Q Z^P+ \eta^{PQ} \gamma_\mu \Zb_P\psi_Q \big).\label{susy}
\end{align}
Here $\eta_{PQ}$ is spinor parametrizing the supersymmetry transformation. It satisfies two constraints: $\eta_{PQ} = - \eta_{QP}$ and $\eta_{PQ} = (\eta^{PQ})^* = \half \e^{PQRD}\eta_{RD}$ leaving $6$ independent complex components. That is, the theory manifestly preserves $12$ supersymmetries.

%%%%%%%%%%%%%%%%%%%%%%%%%%%%%%
%%%%%%%%%%%%%%%%%%%%%%%%%%%%%%

\subsection{Moduli Space}
\label{sect:modulispace}
The moduli space consists of the set of zero-energy field configurations. As usual one sets the fermions to zero, and then looks for $V=0$ states. A sufficient condition for this to occur is
\be
Z^P \Zb_Q Z^R - Z^R \Zb_Q Z^P = 0, \quad \Zb_P Z^Q \Zb_R - \Zb_R Z^Q \Zb_P = 0.\label{zeros1}
\ee
For hermitian matrices, this implies the fields $Z^P$ are diagonal:\footnote{That eqn. \C{zeros1} is a necessary consequence of $\Upsilon_Q^{PR}=0$ is not directly obvious in field theory. However, \C{zeros1} is reasonable when one thinks of ABJM as the IR limit of a intersecting brane construction, as originally developed in \cite{Aharony:2008ug}.}
\be
\langle Z^P\rangle=\diag(z^P_1,\ldots, z^P_N).\label{VEV_1}
\ee
Naively, the field configuration (\ref{VEV_1}) parametrizes a moduli space  $[\IC^4]^N$. However, in ABJM we define global gauge transformations to be part of the gauge group, and as such we need to eliminate gauge equivalent field configurations.\footnote{When the spacetime manifold is non-compact, one is free to interpret global gauge rotations as global symmetries, or as part of the gauge group, the choice is part of the data going into defining the theory. For example, in  \cite{Elitzur:1989nr,Affleck:1989qf}, or in say QED, global gauge transformations are regarded as global symmetries, giving rise to properties such as selection rules. On the other hand, ABJM define global gauge transformations to be  part of the gauge group. This means that in order to determine the moduli space, we need to quotient by them.}
If we restrict to diagonal vev's (\ref{VEV_1}), we only need to worry about two subgroups: the Weyl group, which for $U(N)\times U(N)$ is the symmetric group $S_N$; and the Cartan subgroup which is $(U(1)\times U(1))^N$. The former simply permutes the diagonal elements of (\ref{VEV_1}). As for the latter, it is not hard to see that each scalar field $z_i^P$ is neutral under the diagonal subgroup of $U(1)_D\subset U(1)\times U(1)$ and charged under a baryonic (or axial) subgroup $U(1)_b \subset U(1)\times U(1)$. Thus the classical moduli space is $[\IC^4/U(1)_b]^N/S^N$. However, this is $N(8-1)=7N$ dimensional, which is incompatible with supersymmetry. The resolution is that only a $\IZ_k$ subgroup of global $U(1)_b$ transformations is a symmetry of the quantum theory, as there are semi-classical vacua that carry charge $k$ under the $U(1)_b$. These vacua restrict the global extension of $U(1)_b$ to $\IZ_k$ and therefore the moduli space of the full quantum theory is  $[\IC^4/\IZ_k]^N/S^N$.

Monopole-instanton configurations effect transitions between these vacua. Chern-Simons matter theories have been long known to have monopole-instanton configurations \cite{Lee:1991ge}. In a bifundamental theory such as ABJM, the off-diagonal nature of the Chern-Simons term couples electric $U(1)_b$ and magnetic $U(1)_D$ gauge potentials. Thus, a monopole-instanton with field strength in $U(1)_D$  is also charged under the $U(1)_b$. The saddle point field configuration in the path integral is then charged as the Chern-Simons term transforms under $U(1)_b$ rotations. Denoting our spacetime by $\Sigma_3$ with an asymptotic boundary $\del\Sigma_3$, then the Chern-Simons form transforms under a  $U(1)_b$ rotation $A_b \rightarrow A_b + \del \theta$ as
\be
S_{CS} \to S_{CS} + \frac{k  }{4\pi} \int_{\del\Sigma_3} \theta f_D = S_{CS} +  \theta p k,\label{eq:csshift}
\ee
where $f_D$ is the gauge invariant field strength that carries magnetic charge $\int_{\del\Sigma_3} f_D = 4\pi p$, where $p\in\IZ$ is the monopole-instanton number. We demand the partition function be invariant under gauge transformations, and this is only the case if $\theta = 2\pi/k$. Otherwise, the monopole-instanton vacua are projected out, leading to the mismatch in dimension of the moduli space mentioned above.

%%%%%%%%%%%%%%%%%%%%%%%%%%%%%%
%%%%%%%%%%%%%%%%%%%%%%%%%%%%%%

\subsection{Excitations on the Higgs Branch}
\label{sect:excitations}
Consider now the small excitation spectrum around a generic point on the Higgs branch. Let us recall the usual intuition for D-branes. For $N$ near-coincident D-branes, the low-energy effective field theory describing the dynamics of the system is given by maximally supersymmetric Yang-Mills (SYM). There are scalar fields $X^I$, where $I$ is an R-symmetry label, and the fields transform in the adjoint of the $\U(N)$ gauge group. There is a potential of the form
\be
V \sim -g_{YM}^2 \int \tr [X^I,X^J]^2.
\ee
The minimum occurs when $[X^I,X^J]=0$ implying the scalar fields are mutually commuting and hence  diagonal up to gauge transformations
\be
X^I = \diag(x^I_1,\ldots,x^I_N).
\ee
Taking into account global gauge transformations, the moduli space is $[\RR^d]^N/S_N$ where $d$ is the number of dimensions transverse to the branes. At a generic point in this moduli space, all of the off-diagonal scalar excitations are massive and the unbroken gauge symmetry is $\U(1)^N$. Expanding the potential $V$ about this point in the moduli space, one finds the off-diagonal excitations have a mass
\be
m^2 \sim (x^I_i - x^I_j)^2.\label{dmass}
\ee
The $x_i^I$ are interpreted as parametrizing the location of the $N$ D-branes in the $d$ dimensional transverse space. The quanta of the $X^I$ are open strings connecting the branes.  If $r$ of the $x^I_i$ are equal, corresponding to $r$ D3-branes coinciding, there is an $r\times r$ matrix worth of scalars that become massless. The corresponding W-bosons in the same supermultiplet also become massless, and there is an enhanced $\U(r)$ gauge symmetry. In terms of open strings, the open string excitations connecting the $r$ D3-branes have vanishing length, and hence are massless. From the structure of the potential, it is clear that the enhanced gauge symmetry occurs if and only if the D-branes coincide.

Now let us perform the analogous computation in the M2-brane theory. For simplicity we will from now on restrict to $N=2$ M2-branes, so that our gauge group is $U(2)\times U(2)$. The Higgs branch is  parametrized by
 \bea
\vev{Z^P} = \begin{pmatrix} z^P & 0 \\
0 & w^P \end{pmatrix},\label{VEV_2}
\eea
where the position of the two M2-branes in $\IC^4$ is labelled by $z^P$ and $w^P$. Expanding in small fluctuations about \eqref{VEV_2}
\be
Z^P = \vev{Z^P} + \delta Z^P,\label{fluct1}
\ee
the fluctuations orthogonal\footnote{Fluctuations projected along the VEV,  $\delta Z^p \cdot \langle Z^p \rangle$, are gauge.} to the VEV are described by
\bea
V &=& \frac{4\pi^2}{k^2} \left[ |z^Q \bz_R - w^Q \bw_R |^2 + |z^Q w^R - z^R w^Q|^2\right]\big|\delta Z_{ij}^P\big|^2.\label{pot_1}
\eea
The mass of the off-diagonal modes is then given by
\bea
m^2 = \frac{4\pi^2}{k^2} \left[(|z^P|^2 + |w^P|^2)^2 - 4|z^P \bw_P|^2\right].\label{m2mass_1}
\eea
This formula has a remarkable property quite different from its D-brane cousin \eqref{dmass}, most easily seen if we specialize to the simple scenario where the two M2-branes are separated in a single complex plane i.e. $z^P = w^P= 0$ for $P=2,3,4$. In that case, the mass goes like
\be
m^2 \simeq \big( |z^1|^2 - |w^1|^2\big)^2.\label{m2mass_2}
\ee
This implies there are massless off-diagonal scalar excitations whenever the two M2-branes are at the same radius from the origin, but not necessarily coincident. By taking an arbitrarily large radius, the M2-branes can be separated by an arbitrarily large distance. This behavior differs dramatically from the usual intuition from D-brane theories.

The massless excitations in \eqref{m2mass_2} are not flat directions due to a quartic term in the potential. The quartic term goes like
$$
\frac{4\pi^2}{k^2}(|z^1|^2 + |w^1|^2)|\delta Z^P_{ij}|^4 ,\qquad {\rm for~}P=2,3,4,
$$
which is always non-zero away from the origin. This makes one think that quantum corrections may play an important role.

There are also W-bosons becoming massless along this locus.
As before, consider the special example of $z^P=w^P=0$ for $P=2,3,4$. In that case, there are two W-bosons whose mass goes like (\ref{m2mass_2}). They are given by
\bea
W_1 &=& A_{b\,2}\sin\frac{\theta}{2} - A_{D\,1} \cos \frac{\theta}{2}, \cr
W_2 &=&-A_{b\,1}\sin\frac{\theta}{2} + A_{D\,2} \cos \frac{\theta}{2}.\label{wboson}
\eea
where $A_{b\,\mu} = A_{(1)\,\mu} -  A_{(2)\,\mu}$ and diagonal $A_{D\,\mu} = A_{(1)\,\mu} + A_{(2)\,\mu}$ and we have picked out $\mu=1,2$ components of the baryonic and diagonal gauge fields. The angle $\theta$ is the separation of the M2-branes along the circle of radius $r$ in $Z^1$. Although there are massless W-bosons, for $\theta\ne 0$ there is no enhanced gauge symmetry---the corresponding generators do not close to form a subgroup. At  $\theta=0$, when the branes are coincident but translated from the origin, there is an enhanced gauge symmetry, the diagonal subgroup $\U(2)_D$. When the M2-branes are at the origin there is a further symmetry enhancement to $\U(2)\times\U(2)$. Finally, the excitations are BPS in the same ways as the D-brane excitations discussed around \C{dmass}. With the amount of supersymmetry in our theory, we do not expect the modes to be lifted by any perturbative corrections, even though they are not flat directions.

A way to understand the classical massless excitations is via the action of the $U(1)_b$ gauge symmetry on the vacuum. Suppose the M2-branes are coincident but not at the origin. Then there is an enhanced $U(2)$ gauge symmetry together with the massless off-diagonal scalars. The $\U(1)_b^2$ gauge symmetry acts on the scalar vevs via
$$
z^P \rightarrow e^{i\theta_1} z^P, \qquad w^P \rightarrow e^{-i\theta_2}w^P
$$
separating the M2-branes along a circle in the transverse $\IC^4$. As this is a symmetry of the classical Lagrangian, the fields that are massless when the branes coincide remain massless throughout the gauge orbit. In the quantum theory, the $U(1)_b$ gauge symmetry is broken  down to $\Z_k$ by monopole--instantons. Consequently, when all the quantum corrections are taken into account, we expect the $U(1)_b$ degeneracy and the associated massless excitations to be lifted.

Let us turn now to the dynamics of the light fields $z^P,w^P$ at a generic point on the Coulomb branch where the off-diagonal modes are massive. In fact, let us simplify life a wee bit by considering just the dynamics of $z^P$; the dynamics of $w^P$ will follow analogously. The field $z^P$ is governed by an effective action
\be
S_{eff} = - \int_\Sigma |D_\mu z^P|^2 + \frac{k}{4\pi} a_b \w f_D + \ldots,
\label{effective1}
\ee
with terms omitted of the order the Higgs mass. Even though $a_b$ has been Higgsed, and enjoys a mass via a Chern-Simons-Higgs mechanism, we have not integrated it out, as it still has a role to play. The field $z^P$ couples only to the baryonic $U(1)_b$ gauge group and is neutral under the $U(1)_D$ as illustrated by the covariant derivative:
\be
D_\mu z^P = \del_\mu z^P - i a_{b\,\mu} z^P.
\ee
Indeed, the $U(1)_D$ gauge field appears in $S_{eff}$ only via its field strength $f_D$. We can dualize it into a scalar by introducing a lagrange multiplier imposing the Bianchi identity for $f_D$
\be
S_\tau = \frac{1}{4\pi} \int_\Sigma \tau  d f_D.\label{bianchi1}
\ee
The equation of motion for $\tau$ enforces the Bianchi identity $df_D = 0$. If $\Sigma$ has a boundary, then in the presence of monopoles with field strength in $f_D$ there is a periodicity constraint on the zero-mode of $\tau$. A monopole localized in $\Sigma$ will have its charge quantised $\int_{\del \Sigma} f_D = 4\pi n$. Then, $S_\tau$ pulls back to an integral on the boundary $\del \Sigma$ and is equivalent to adding an operator to correlation functions of the form
\be
\cO = e^{iS_{\tau}}  = e^{i\tau n}.\label{bianchi2}
\ee
This implies the zero-mode of $\tau$ is periodic $\tau \sim \tau + 2\pi$. As $f_D$ is now unconstrained by the Bianchi identity, we can integrate it out by imposing its equation of motion
\be
a_b = \frac{1}{k} d\tau.\label{AxialEOM}
 \ee
Under a global $\U(1)_b$ transformation $z^P\rightarrow z^Pe^{i\theta}$, the relation \C{AxialEOM} implies $\tau \sim \tau  + k\theta$. The analysis for the $w^P$ scalar field follows in the same way, implying we end up with two scalars dual to the $U(1)_b^2$ photons. Our motivation for introducing the dual photons is they are needed to construct local gauge invariant monopole-instanton vertex operators in the effective action.

%%%%%%%%%%%%%%%%%%%%%%%%%%%%%%
%%%%%%%%%%%%%%%%%%%%%%%%%%%%%%

\subsection{Light states in related M2-branes Theories}
The appearance of massless excitations at special points in the moduli space was also noticed by \cite{Lambert:2008et} in the context of the Bagger-Lambert-Gustavsson (BLG) model. In that case, excitations became massless when the M2-branes were collinear with the orbifold fixed point. At the level of the classical Lagrangian, it is straightforward to map the BLG theory to the $\SU(2)\times\SU(2)$ ABJM theory by a field redefinition (see \cite{Lambert:2010ji} for a related discussion). Using the explicit field redefinition, we show in appendix \ref{app:BLG} how the singular locus noted in \cite{Lambert:2008et} maps to the ABJM singular locus discussed above. In particular, one expects that non-perturbative corrections in BLG are likely to play a similar role to the discussion presented here for ABJM. In appendix \ref{app:cy4} we show how a singular locus appears in more general ABJM-like theories, for example those probing toric Calabi-Yau four-folds. It is clear that whatever physics resolves the singular locus and associated light states in ABJM will apply in these associated contexts.

%%%%%%%%%%%%%%%%%%%%%%%%%%%%%%
%%%%%%%%%%%%%%%%%%%%%%%%%%%%%%

\section{Monopole--Instantons}
The effective action of the light modes $z^P,w^P$ at a generic point on the Coulomb branch receives quantum corrections. Supersymmetry forbids any non-trivial perturbative corrections, leaving one to consider non-perturbative corrections. Non-perturbative corrections that we consider here arise in the form of instanton corrections: finite action Euclidean solutions of the classical equations of motion, which preserve some amount of supersymmetry. Instantons in three-dimensions arise as the dimensional reduction of monopoles in $3+1$ dimensions along the time direction. These field configurations are classified by a topological invariant and form a saddle point about which we perform the path integral. Constructing these instanton and evaluating their semi-classical contribution to the effective action is the subject of this section.

%%%%%%%%%%%%%%%%%%%%%%%%%%%%%%
%%%%%%%%%%%%%%%%%%%%%%%%%%%%%%

\subsection{Constructing the monopole--instanton solution}
We construct a solution to the Euclidean equations of motion, largely following and slightly improving on the analysis in \cite{Hosomichi:2008ip}. The action \C{action1} after a Wick rotation $t = -i\tau$ becomes
$$
-S_E = S_{KE} + S_{\rm int} + S_V + iS_{CS},
$$
The gauge field  equations of motion in Euclidean signature are given by
\bea
 \frac{k}{2\pi}\star F_{(1)} &=&  (D Z^P)\Zb_P -  Z^P (D\Zb_P),\cr
\frac{k}{2\pi} \star  F_{(2)} &=&   (D\Zb_P) Z^P -  \Zb_P (DZ^P) .\label{CSEOM1}
\eea
The lack of a Maxwell term for $A_{(i)}$ means the gauge field has no independent dynamics---its behaviour is completely tied to the dynamics of the matter fields.  For simplicity we assume the solution is confined to a single complex plane, so that $Z^P =0$ for $P=2,3,4$ and label the remaining field $Z^1 = Z$. We wish to preserve some supersymmetry, which from \C{susy} gives rise to a BPS condition:
\be
DZ = 0.\label{BPS}
\ee
BPS instanton solutions of \C{CSEOM1} will in general be complex,  meaning $Z^\dagger \ne \Zb$. This is a generic property of Chern-Simons matter theories, as well as more general theories in which the gauge field has a term linear in time derivatives (e.g. \cite{Freed:1990uw}). Physically, we interpret the instanton as a tunneling solution, taking physical vacua to physically inequivalent vacua. The vacua obey the reality constraint $Z^\dagger = \Zb$ meaning the instanton solution, though complex in the interior of $\Sigma$, must be real on the boundary $\del \Sigma$. We will address this issue later.

The equations of motion \C{CSEOM1} together with the BPS condition \C{BPS} give
\bea
 \frac{k}{2\pi}\star F_{(1)} &=&  - iD (Z \Zb) ,\cr
\frac{k}{2\pi} \star  F_{(2)} &=&  i D(\Zb Z) .\label{CSEOM2}
\eea
These equations resemble the usual Bogomol'nyi equation describing a `t~Hooft--Polyakov monopole, which together with the knowledge that ABJM on the Coulomb branch can be rewritten as a Yang--Mills theory \cite{Mukhi:2011jp}, leads one to search for `t Hooft--Polyakov like solutions. To that end, we first make the ansatz $A_{(1)}=A_{(2)}$. This has several justifications. Asymptotically, physical considerations imply $Z$ and $\Zb$ become diagonal in order to be vacuum states. This implies $F_{b}$ vanishes asymptotically, and as such, the BPS solution only has a non-trivial field strength in the diagonal subgroup. Further, only the diagonal generators close to form a group; the non-abelian baryonic (or axial) generators do not close  to form a group. Finally, the abelian diagonal subgroup $\U(1)_D^2$ is the only subgroup of $\U(2)\times\U(2)$ that is unbroken in the vacuum. The remaining components are spontaneously broken, and hence cannot carry the monopole field strength.

Now rewrite (\ref{CSEOM2}) in terms of the diagonal and baryonic bases:
\bea
 \frac{k}{2\pi}\star F_{D} &=& j_{b} ,\cr
\frac{k}{2\pi} \star  F_{b} &=& 0,\label{CSEOM3b}
\eea
where $j_{b} = j_{(1)}-j_{(2)}$ is the baryonic matter current. The general form of the solution to \eqref{CSEOM3b} together with the BPS condition $DZ=0$ is of the form
\bea
Z = L(aI_2)M^{-1}, \quad \Zb = M(b \Phi + c I_2)L^{-1},\label{monopole_ansatz}
\eea
where the equations of motion amount to $\Phi$ satisfying the Bogomol'nyi equation
\be
\star F_D = m\, D\Phi, \quad m = \pi ab /k.\label{eqn:bog_2}
\ee
An explicit solution of the Bogomol'nyi equation involves a gauge choice. As a first attempt, we choose $L=M=1$ and solve the Bogomol'nyi equation using the `t Hooft--Polyakov monopole. We describe the field configuration of the monopole in Hedgehog gauge, and denote the Hedgehog gauge field configuration by $\cA$. Then, the solution to \C{eqn:bog_2} is
\bea
\Phi = \frac{\hat{r}^\mu \sigma^\mu}{2rm} \left( m r \coth m r - 1\right), \qquad A_D = \cA= \eps_{\mu\nu\rho} \frac{\sigma^\mu \hat{r}^\nu}{2r}\left(1-\frac{m r}{\sinh m r}\right) dx^\rho,   \label{eqn:bog_1}
\eea
where $r=|x|$. The parameter $m$ is the mass scale of the monopole; it defines the size of the core of the monopole in which the non-abelian gauge fields become excited. Outside the core, $r\gg m$, the fields behave as
\be
\Phi \sim \half\hat{r}^\mu\sigma^\mu, \quad  \cA = \eps_{\mu\nu\rho} \frac{\sigma^\mu \hat{r}^\nu}{2r}dx^\rho,
\ee
and
\be
\cF \sim \frac{d\hat{r}}{r^2} d\theta\w d\phi.
\ee
The abelian component of the monopole is long-ranged, being only power-law suppressed.  With these conventions the enclosed magnetic flux is
\be
\int_{S^2_\infty} \cF = 4\pi\ .
\ee
The constants $a,b,c$ in \C{monopole_ansatz} are determined by boundary conditions. As expected the solution \C{monopole_ansatz} is complex, even asymptotically. We can patch this up by a judicious choice of $L,M$ in \C{monopole_ansatz}
\be
L = f(x)e^{\Lambda(x) \Phi(x)},\quad M = f(x)e^{-\Lambda(x) \Phi(x)},\label{eqn:L_transf}
 \ee
where $\Lambda(x)$ is a function designed so that $Z,\Zb$ have real boundary conditions, while the role of $f(x)$ is to implement a discrete $SU(2)$ Weyl transformation at the beginning of time so the monopole has the correct tunnelling interpretation. The gauge fields are related to $\cA$ in  \C{eqn:bog_1} by
\bea
A_{(1)} = L \cA L\inv - i dL  L\inv, \quad A_{(2)} = M \cA M\inv - i dM M\inv.\label{eqn:gauge_2}
\eea
Do not be fooled: this is not necessarily a gauge transformation, as $L,M$ need not be unitary. Nonetheless, as \C{monopole_ansatz}, \C{eqn:gauge_2} take the same form as \C{eqn:gauge1}, we can regard it as a field redefinition with the attribute that the supersymmetry conditions, equations of motion and solution transform covariantly, thereby mapping BPS solutions to BPS solutions in a 1-1 fashion. Furthermore, the similarity to a gauge transformation means the measure in the path integral is invariant under this transformation.

We still need to specify the functions $f(x),\Lambda(x)$. The function $f(x)$ is given by
\be
f(x) = \frac{1}{2}(1-\tanh \tau)\sigma_1 = e^{\frac{i\pi}{4}(1-\tanh\tau) \sigma_1}\label{eq:f},
\ee
while $\Lambda(x)$, as well as $a,b,c$, are fixed by first specifying the in and out vacua
\bea
\vev{Z_{i}} &=& \bpm z_{i} & 0 \\ 0 & w_{i} \epm, \label{initial}\\
\vev{Z_{f}} &=& \bpm z_{f} & 0 \\ 0 & w_{f} \epm.\label{final}
\eea
and then looking at the boundary conditions are the ending $\tau\rightarrow \infty$ and beginning of time $\tau\rightarrow +\infty$.

\ben
\item {\sl End of time $\tau \rightarrow \infty$: }
In this case $\Lambda(x)\rightarrow \Lambda_+$ and $\Phi \rightarrow \frac{1}{2} \sigma^3$ giving
$e^{2\Lambda\Phi} \sim \diag(e^{\Lambda_+}, e^{-\Lambda_+})$. Plugging into \C{monopole_ansatz} and comparing with \eqref{final} we can fix $\Lambda_+$ and the constants $a,b,c$ in the ansatz \C{monopole_ansatz}:
\bea
a = \sqrt{z_f w_f}, \quad e^{\Lambda_+ } = \sqrt{z_f/w_f}, \quad b = \frac{|z_f|^2-|w_f|^2}{\sqrt{z_fw_f}}, \quad c = \frac{|z_f|^2+|w_f|^2}{2\sqrt{z_fw_f}}.\label{monopolecoefficients}
\eea

\item {\sl Beginning of time $\tau\rightarrow-\infty$:}
In this case $\Lambda(x)\rightarrow \Lambda_-$, and $\Phi \rightarrow -\half \sigma^3$ giving $e^{2\Lambda\Phi} \sim \diag(e^{-\Lambda_-}, e^{\Lambda_-})$.
The constants $a,b,c$ have already been determined, but $\Lambda_-$ has not. Using \eqref{monopole_ansatz} and \eqref{initial} we find
\bea
e^{\Lambda_-} = \sqrt{w_i/z_i},
\label{eq:Lbegin2}
\eea
as well as the consistency conditions
\bea
z_f/z_i &=& \frac{1}{w_f/w_i}, \quad |z_f|=|z_i|,\quad |w_f|=|w_i|,\cr
&\Rightarrow& z_f = e^{-i\theta} z_i, ~~~ w_f = e^{i\theta}w_i.\label{charge_1}
\eea
These conditions%
\footnote{Without $f(x)$ the monopole--instanton would also flip the M2-branes $z\leftrightarrow w$. If as in \cite{Hosomichi:2008ip} we were to take $f(x)=1$, $\Lambda_+=\Lambda_-$ in the unitary gauge, with $z=u_1,w=u_2$ real, then it is not hard to see that the consistency conditions would force $\theta=0$ and $u_1 = u_2$. This solution does not have the interpretation of a tunneling solution, hence our different choice of $f(x),L,M$ and parametrization of the moduli space by complex scalars.} tell us the monopole is transferring $\U(1)_b$ charge consistent with the monopole being sourced by $j_b$ in \eqref{CSEOM3b}.
\een

\noindent
A smooth function $\Lambda(x)$ satisfying these boundary conditions is
\bea
\Lambda(x) &=& \left[ \frac{1}{4} (1+ \tanh \tau) \log\left( \frac{z_f}{w_f}\right)\, + \frac{1}{4} (1- \tanh \tau) \log \left(\frac{z_i}{w_i}\right)\right].\label{eqn:lambda_1}
\eea
To summarise, the monopole solution is
\bea
Z &=& \sqrt{z_f w_f} \,\,f(x) e^{2\Lambda(x)\Phi(x)}f(x), \cr
\bZ &=& f(x)\frac{e^{-2\Lambda(x)\Phi(x)}}{\sqrt{z_fw_f}} \Big((|z_f|^2-|w_f|^2)\Phi(x) + (|z_f|^2+|w_f|^2)I_2\Big)f(x),\label{monopole_1}
\eea
with $f(x)$ given by \C{eq:f} and $\Lambda(x)$ given by \eqref{eqn:lambda_1}. The monopole acts on states by
\bea
\bpm z_i & 0 \\ 0 & w_i \epm_{\tau=-\infty} \longrightarrow \bpm z_f & 0 \\ 0 & w_f\epm_{\tau=\infty}=\bpm z_i e^{-i\theta}  & 0 \\ 0 & w_i e^{i\theta}  \epm.  \label{monopole_action}
\eea

%%%%%%%%%%%%%%%%%%%%%%%%%%%%%%
%%%%%%%%%%%%%%%%%%%%%%%%%%%%%%

\subsection{Evaluation of the classical action}
It will be useful for us later to evaluate the classical action in this background. There are three terms to consider:
\be
- S_E = S_{KE} + S_V + iS_{CS} .
\ee
The term $S_{KE}$ vanishes due to the BPS condition $DZ=0$, while the potential term $S_V$ vanishes as we are on the moduli space. This leaves the Chern-Simons action $S_{CS}$. As the field redefinition \eqref{eqn:gauge_2} has the same form as a gauge transformation, it is straightforward to evaluate $S_{CS}$ using \eqref{eqn:gauge1}-\eqref{cstransformation}.
\begin{equation}
\begin{split}
S_{CS} ~=~& \frac{k}{4\pi}  \int_\Sigma \omega_{CS}(A_{(1)}) - \omega_{CS}(A_{(2)}),\cr
~=~& \frac{-ik}{4\pi} \int_\Sigma  d \big[\tr\big(\cA L\inv dL\big) - \tr\big(\cA M\inv dM\big)\big] - \frac{k}{12\pi} \int_\Sigma \tr (L\inv dL)^3   +  \tr (M\inv dM)^3.
\end{split}
\end{equation}
The first line restricts to the boundary $\del\Sigma$, while the last line evaluates to the winding numbers of $L,M$, which cancels. We evaluate the first line on the asymptotic two-sphere $S^2_{\infty}$, which truncates us to the linear approximation.
\be
-S_E = \frac{k}{2\pi} \int_{S^2_\infty} f(x)(\Lambda\Phi^a)f(x) \mathcal{F}_{a},
\ee
where we have used the following normalization on the Lie algebra generators: $\tr\, T^a T^b = \half \delta^{ab}$. Split the integral into a sum over the upper hemisphere $S^2_+$ and lower hemisphere $S^2_-$. By spherical symmetry and the property of $f(x)$:
$$
\int_{S^2_\pm} f(x)\Lambda(x)\Phi^a(x) f(x) \mathcal{F}_{a}  =  \pm2\pi \Lambda_{\pm},
$$
which after using \eqref{eq:Lbegin2} we find
\bea
-S_E &=& \frac{k}{2\pi} \Big[ \int_{S^2_+} \Lambda_+(x)\Phi^r(x) \mathcal{F}_{r} - \int_{S^2_-} \Lambda_-(x)\Phi^r(x) \mathcal{F}_{r}  \Big]\cr
&=& \frac{k}{2} \left(\log \frac{z_f}{w_f} + \log \frac{z_i}{w_i}\right),
\eea
where $\cF_r =\hat r^a \cF_a$. Thus, the monopole-instanton contributes
\be
e^{-S_E} =  \left( \frac{z_i}{w_i}e^{-i\theta}\right)^{k},
\label{classicalaction1}
\ee
to the path integral.

%%%%%%%%%%%%%%%%%%%%%%%%%%%%%%
%%%%%%%%%%%%%%%%%%%%%%%%%%%%%%

\section{Monopole--Instantons and light excitations}
Previously we observed the small-fluctuation analysis around the vacuum exhibited a pole when two M2-branes are separated by a $U(1)_b$ transformation. The pole represented the off-diagonal modes becoming light. Can we use the monopole-instanton solution to determine the physics of these off-diagonal light modes?

As a warm-up, we can study the excitation spectrum in the instanton background by a simple generalization of the analysis performed above for the vacuum. We expand the scalar fields as
\be
\begin{split}
Z^P &= \delta^{P1} \langle Z \rangle + \delta Z^P =  \sqrt{z_f w_f} L^2 + L \delta \cZ^PL, \cr
\bZ^P &=\delta^{P1} \langle \Zb \rangle + \delta \Zb^P \cr
&= \frac{e^{-2\Lambda(x)\Phi(x)}}{\sqrt{z_fw_f}} \Big((|z_f|^2-|w_f|^2)\Phi(x) + (|z_f|^2+|w_f|^2)I_2\Big) + L^{-1} \delta \overline{\cZ}_PL^{-1},
\end{split}
\label{small_1}
\ee
where $\vev{Z},\vev{\Zb}$ denote the monopole-instanton background \eqref{monopole_1} and $\delta Z, \delta \Zb$ are the small fluctuations. It is convenient to first perform the analysis for the case where the monopole-instanton is confined to a single complex plane, though we will later expand the analysis to an $\SU(4)$ covariant expression. The off-diagonal excitations intertwine with the transformation $L$ defined in \eqref{eqn:L_transf}, $L \delta \cZ L = \delta Z$, and we need to keep in mind the fields $Z^P$ satisfy the reality constraint $Z^{P\,\dagger} = \Zb^P$ asymptotically.

Now expand the scalar potential \eqref{pot} about this background:
\bea
\delta \Upsilon^{PR}_Q &=&  (|z_f|^2-|w_f|^2) \, \delta_{Q1} \left(\delta^{R1} [\delta Z^P,\Phi] - \delta^{P1} [\delta Z^R,\Phi]\right),\cr
\delta \overline{\Upsilon}_{PR}^Q &=& (|z_f|^2-|w_f|^2) \, \delta^{Q1} \left(\delta_{R1} [\delta \bZ_P,\Phi] - \delta_{P1} [\delta \bZ_R,\Phi]\right).
\eea
Hence,
\be
V = \frac{4\pi^2}{3k^2} (|z_f|^2-|w_f|^2)^2 \tr [\delta Z^P,\Phi][\delta \Zb_P,\Phi],
\ee
As the vacuum is approached at $\tau\rightarrow\pm\infty$, then $\Phi \rightarrow \pm \half \sigma^3$ and one finds a multiplet of massive states with mass \eqref{m2mass_2}.
Hence, just as for the vacuum in section \ref{sect:excitations}, there is a locus of singularities when $|z|=|w|$. This phenomenon persists for higher order monopole-instanton numbers thanks to the putative $U(1)_b$ symmetry, implying that order-by-order in the monopole-instanton expansion, the off-diagonal modes remain massless.

Indeed, in the low-energy effective action on the Higgs branch, supersymmetry prohibits the generation of a mass term via quantum corrections. In a derivative expansion, the lowest order term that is quantum corrected is an 8-fermion coupling, or equivalently, a four-derivative coupling of the scalars. These couplings determine the strength with which M2-branes scatter, and are generated by monopole-instantons. Do these couplings have any bearing on the massless off-diagonal modes along the $U(1)_b$ locus?

As explored by \cite{Dorey:1997tr,Polchinski:1997pz,Paban:1998mp}, $N=8$, $d=3$ SYM possesses an analogous $8$-fermion coupling that contributes to the strength of D2-brane scattering. Order by order in the monopole-instanton expansion, the coupling exhibits a pole from light states only when the D2-brane coincide in $\RR^7$. However, when viewed from the point of view of M-theory, this is really an M2-brane scattering process in $\RR^7\times S^1$. Is this fact visible in $N=8$ $d=3$ SYM? Firstly, the M-theory circle coordinate is not manifest in the classical Lagrangian: it appears via the dual photon for the $U(1)$ gauge group when on the Higgs branch. Secondly, at at any given order in the monopole-instanton expansion, the theory exhibits a degeneracy along the $S^1$. This is in contrast to our M-theory expectations, in which we expect the branes to be localized in $\RR^7\times S^1$, so how does one reconcile this with what is seen in the gauge theory? A clue comes from studying the 11-dimensional supergravity dual, where Fourier expanding the M2-brane interaction potential along the M-theory circle direction corresponds to a semi-classical expansion in monopole--instantons in the gauge theory. At each order in the Fourier expansion, the potential is smeared in the $U(1)$ direction, and it is only once all the terms are summed that the degeneracy is removed. In the gauge theory, this means the degeneracy in the $U(1)$ direction of field space is removed only once all the monopole--instantons are summed.

There are some striking analogies with what we have seen thus far in ABJM. The theory exhibits a  degeneracy along the $U(1)_b$ direction, implying the M2-branes are smeared in this direction. At each order in the monopole-instanton expansion the theory preserves this degeneracy. Unlike SYM this gives rise to new massless states, even when the M2-branes are not coincident. This seems to be a consequence of a direction in the moduli space being tied to a classical gauge symmetry. Nonetheless, the analogy with SYM leads one to suspect that the degeneracy will be lifted once the monopole--instantons are summed.

One can give evidence for this effect by considering the supergravity dual of the gauge theory. This is given by a pair of M2-branes moving in the background $\IC^4/\Z_k$, and by treating one M2-brane as a background source, one can compute what the low-energy effective action is on the probe brane.
%We give the details of this calculation in both SYM and ABJM in Appendix \ref{app:sugra}, and quote the results here.
The first non-trivial interaction term in the DBI expansion is
\be
\frac{1}{8\pi^2} \int d^3 \sigma F(\vec{z},\vec{w}) |\vec{\dot{z}}^4| + \ldots, \quad {\rm where} \quad F(\vec{z},\vec{w})  = \sum_{l=0}^{k-1} \frac{1}{|\vec{z} - \vec{w} e^{-2\pi i l/k}|^6},
\ee
and we are summing over images of the orbifold action, so that we can work in the $\IC^4$ cover. Writing $\IC^4$ as a cone over $S^7$, the $S^7$ admits a Hopf fibration with base $\IC \PP^3$. The function $F(\vec{z},\vec{w})$ then admits a Fourier expansion $F = \sum_{p=-\infty}^{p=+\infty} f_p(z,w) e^{ip\theta}$ where $\theta$ parametrizes the $S^1$ fibre of the Hopf fibration. This $\theta$ is interpreted as the M-theory circle, and from the ABJM point of view it is the $\U(1)_b$ direction along which the monopole carries electric charge. Details of this expansion will be given below.  The Fourier coefficients are interpreted as monopole-instanton corrections in ABJM with instanton number given by $p$. The interaction term is expanded as
\be
 F(\vec{z},\vec{w}) = \sum_{p=-\infty}^{p=\infty}  f_{p}(\vec{z},\vec{w}) e^{ikp\theta},\label{Fourier4a}
\ee
where in the case where $z^{2,3,4}=w^{2,3,4}=0$, the Fourier coefficients take the form
\begin{equation}
\begin{split}
f_{p}(z,w) ~=~& \frac{ 4\pi^3}{k^4 m^5} \left(\frac{|z|}{|w|}\right)^{k|p|}   \Big\{8\pi^2(|\vec{z}|^4 + 4|z|^2|w|^2 +  |\vec{w}|^4) +  \cr
 & \quad+ 6\pi|p|k^2(|z|^2 + |w|^2)m  +  p^2k^4 m^2\Big\}.
\end{split}
\label{Fourier3b}
\end{equation}
Without loss of generality we have assumed $|z|<|w|$.

Following \cite{Polchinski:1997pz,Dorey:1997ij} we interpret the Fourier coefficient $f_{p}$ as the coefficient of the corresponding vertex in the gauge theory generated by the monopole-instanton with charge $kp$.  That main point is that for each instanton number $p$, the coefficient, $f_{p}$, is manifestly $\SU(4)$ invariant and smeared in the $\U(1)_b$. However, when all of the coefficients are summed, as is done in \eqref{Fourier4a}, the resulting expression $F(\vec{z},\vec{w})$ is manifestly $\SO(8)$ invariant modulo the $Z_k$ quotient, with the smearing in the $\U(1)_b$ direction removed. From the point of view of the gauge theory, this implies once the instantons are summed, the $\U(1)_b$ degeneracy in the classical theory will be lifted. It is then natural to conclude that the corresponding off-diagonal modes, which were generated by the putative $U(1)_b$ symmetry, are in fact in some sense massive in the full quantum theory. This is so, even though at each order in the instanton sum, the result looks smeared in the $\U(1)_b$ direction and the corresponding off-diagonal modes are massless.

However, seeing this explicitly in ABJM is tricky. Firstly, the monopole--instantons do not directly generate a mass-term, instead generating a higher-derivative coupling in the effective theory. Relating this to the dynamics of the light off-diagonal excitations is subtle. Secondly, seeing the decoupling of the light off-diagonal excitations requires an all-order instanton calculation, which we do not yet have. Nonetheless, in the remainder of this section we present the calculation for the $p=1$ monopole-instanton contribution to the effective action. The resulting 8-fermion coupling is related by supersymmetry to the coefficient $f_{1}$ written in \eqref{Fourier3b} above. The technology developed here should be extendable to an all-order calculation in the near future.

%%%%%%%%%%%%%%%%%%%%%%%%%%%%%%
%%%%%%%%%%%%%%%%%%%%%%%%%%%%%%

\subsection{Monopole--instanton calculation of $8$-fermion correlator}
The first step is to evaluate the 8-fermion correlator in the microscopic theory using the monopole-instanton background developed above. Local symmetries tightly constrain the types of correlators we can compute. There are 8 fermion zero modes to soak up, and the insertion should be $U(2)\times U(2)$ gauge invariant. In particular, it should be invariant under the global $U(1)_D$, as well as $SU(4)_R$ invariant. This leaves only two choices:
\be
\epsilon_{QRST} \psi_{P\,\alpha} \psib_{ \beta}^QZ^P \psib_{ \gamma}^R Z^S \psib_{ \delta}^T,\quad Z^P \Zb_R \psi_{P\,\alpha} \psib_{ \beta}^Q\psi_{Q\,\gamma} \psib_{ \delta}^R,\label{su4fermions}
\ee
where we have exhibited the $SO(2,1)$ spinor indices $\alpha,\beta,\ldots$ explicitly. Only the first term is relevant if we choose the vev to lie in a single complex plane, and as the monopole--instanton as 8 fermion zero modes, we need the square of this term contracting spinor indices pairwise $(\psi_{P})^2$. Hence, the local symmetries have told us we need to calculate the following correlator in the monopole-instanton background:
\be
\Big\langle\tr\big[\epsilon_{QRST}  \psi_P \psib^QZ^P \psib^R Z^S \psib^T\,\big]^{\,2}\Big\rangle.\label{correlator1}
\ee
The calculation then proceeds by integrating over the non-zero modes, leaving a finite integral over the zero modes of the background:
\beq
\begin{split}
&\big\langle\,  \tr\big[\epsilon_{QRST}  \psi_P \psib^QZ^P \psib^R Z^S \psib^T\,\big]^{\,2} \, \big\rangle = \cr
&\qquad \qquad\quad =\int d \mu_B \int d\mu_F \Dloop e^{-S_{cl}} \tr\big[\epsilon_{QRST}  \psi_P \psib^QZ^P \psib^R Z^S \psib^T\,\big]^{\,2}.
\end{split}
\label{schematic}
\eeq
The zero mode measures are calculated in Appendix \ref{app:zeromode} and are given by \C{eq:bosonicmeasure} and \C{fermionmeasure}:
\be
\begin{split}
\int d\mu_B ~=~& k^2 m^3 \pi^2 d^3 X_{cm} d\phi, \cr
\int d\mu_F ~=~& \frac{1}{4 (2\pi k)^4 (b^2 d)^2 m^2}\int d^8 \eta.
\end{split}
\label{results1}
\ee
Here $X_{cm}$ denotes the monopole-instanton center of mass and $\phi$  parametrizes global $U(1)_D$ gauge rotations under which the monopole carries field strength. The integrand will be independent of this direction, and hence the $\phi$ integral will just give a factor of $2\pi$.  The constant $b$ is determined in \eqref{monopolecoefficients}, while $d$ is fixed in terms of the vevs $\vec{z},\vec{w}$ in \eqref{vevsd1}-\eqref{vevsd2}:
\be
|d|^2 = \frac{\tr Z^2 \bar Z_2}{\tr Z^1 \bar Z_1} = \frac{|z_2|^2 + |w_2|^2}{|z_1|^2 + |w_1|^2}.\label{dconstant}
\ee
As discussed in appendix \ref{app:zeromode}, to perform the zero-mode integrals one needs to put the M2-branes at a more generic point in the moduli space. In particular, this implies $\langle Z^{1,2}\rangle\ne0$ up to the $SU(4)_R$ symmetry. The end result will however be independent of $d$ and so we can take the limit where $\langle Z^2 \rangle = 0$ if we wish, without causing any difficulty.
The 1-loop determinant, denoted by $\Dloop$ is calculated in Appendix \ref{app:1loop} and is given by
$$
\Dloop = 2^{-4}.
$$
The evaluation of the classical action is cf. \C{classicalaction1},
$$
e^{-S_{cl}} = \left(\frac{z}{w}\right)^{k}.
$$
The insertion $\tr\big[\epsilon_{QRST}  \psi_P \psib^QZ^P \psib^R Z^S \psib^T\,\big]^{\,2}$ in \C{schematic} soaks up the zero modes of the instanton \C{fermionzero1}, which in the long--distance limit looks like cf. \C{fermionzero2}:
$$
 \tr\big[\epsilon_{QRST}  \psi_P \psib^QZ^P \psib^R Z^S \psib^T\,\big]^{\,2} = (8\pi k)^4 (b^2 d)^2 (G_{LD}\eta)^8 m^{-4}.
$$
With our explicit choice of vev, we can now put the pieces of the jigsaw together to evaluate the the correlator \C{correlator1}:
\bea
\big\langle\,  \tr(\psi_1 \psib^{\,2} Z^1\psib^{\,3}Z^1\psib^{\,4})^2  \, \big\rangle =  \frac{(2\pi)^3 k^2}{ m^3} \left(\frac{z}{w}\right)^{k} \int d^3 X_{cm} \prod_{i=1}^8 \GLD(X-x_i),\label{correlatorfinal}
\eea
The answer is invariant under all the relevant symmetries, including the $\U(1)_D$, up to the spontaneous breaking by choice of vev.

\subsection{Effective action and a comparison with supergravity}
We now turn to the effective action of ABJM on the Higgs branch, where the action takes the schematic form
\be
S = S_{\text{\it free}} + S_{\text{\it 8-fermion}}  + \ldots
\ee
The omitted terms are the supersymmetric completion of the $8$-fermion term. This term can be attributed to monopole--instantons, and takes the form
\be
S_{\text{\it 8-fermion}} = \int d^3X \sum_{p=1}^{\infty}g_{p}(z^P,w^P,\psi_z^P,\psi_w^P),\label{vertex0}
\ee
where $g_{p}$ contains fermion zero-modes and combinations of the scalar fields. In the language of effective field theory, there is a vertex $g_1$ that reproduces the monopole-instanton correlator \eqref{correlatorfinal}. Are there any independent checks of \eqref{correlatorfinal}? As mentioned in \cite{Hosomichi:2008ip}, one can appeal to the supergravity description of the system. In the supergravity limit, the four-derivative interaction term that is most easily calculated is the $v^4$ coupling mentioned above, which is in the supersymmetric completion of the 8--fermion vertex. Unfortunately, it is not easy to directly relate the two thanks to the complicated nature of the supersymmetry transformations. One would ideally like a more direct check. In the context $N=8$ $SU(2)$ SYM, \cite{Paban:1998mp} were able to constrain the form of the $8$-fermion vertex $g_p$ using various properties of the supersymmetry algebra, and infer the eight fermion interaction in M-theory. Although the microscopic theories are distinct, the IR fixed points are related and so we might hope to compare the result in \cite{Paban:1998mp} with our result in \eqref{correlatorfinal}.

First we briefly describe the result in \cite{Paban:1998mp}. To do so we need to briefly introduce some notation. Denote the $SU(2)$ adjoint valued scalar field by $\wt \phi^i$, with $i=1,\ldots,7$ the $SO(7)$ R-symmetry indices. As we are looking at the effective theory on the Higgs branch, the gauge theory becomes abelian in the IR limit  with the light field defined as $\phi^i=\tr \sigma_3 \wt \phi^i$. A similar definition applies to the light fermions,  which are denoted by $\psi_{\alpha a}$. These transform in the $(\rep{2},\rep{8})$ of the $SO(2,1)\times{\rm Spin}(7)$ symmetry group, where $\alpha=1,2$ are the SO(2,1) spinor indices and $a=1,\dots,8$ the spinor R-symmetry indices. There is also a dual photon, denoted by $\phi_8$. At the IR fixed point, \cite{Paban:1998mp} showed there was an 8-fermion vertex given by (up to an overall coefficient):
\beq
S_{\text{\it 8-fermion}} = \int d^3 x \frac{1}{r^{10}} \big[a_1(\psi_1 \gamma^{pr} \psi_1)^2(\psi_2 \gamma^{qr} \psi_2)^2 + a_2(\psi_1 \gamma^{pr} \psi_1 \psi_2 \gamma^{pr} \psi_2)^2 \big],\label{vertex1}
\eeq
where $r^2 = \phi_1^2 + \dots \phi_8^2$ and $\gamma^r = \phi^i \gamma^i / r$.  Here the subscripts $1,2$ on the fermions are explicit $SO(2,1)$ spinor indices.

Our task is to now relate this result to ABJM by changing coordinates, orbifolding and Fourier transforming in an appropriate $U(1)$ direction. The first coefficient of this Fourier transform is interpreted as the $p=1$ instanton coefficient.
%First we change coordinates.  Rewrite $\IR^8$ as $\IC^4$ with the two membrane locations denoted by $SU(4)$ vectors $\vec{z},\vec{w}$. $\IC^4$ can be written as a cone over $S^7$, which in turn can be written as a $U(1)$ fibration over a base $\IC\IP^3$. The instanton calculation is naturally written in a coordinate system compatible with this structure. To that end introduce a field $\rho$ whose expectation value is $\langle \rho^2 \rangle = m_{12}$. We can think of this as being a natural radial coordinate. There are also classically massless fields corresponding to the $U(1)_b$ fiber direction, denoted by $\theta$ and  angular rotations of the $\IC\IP^3$ denoted by $\vec{\omega}$. These explicit forms of these fields will not play a crucial role in the following. The $SU(4)_R$ symmetry rotates the $\vec{\omega}$ fields, transforming in the $\rep{6}$, while $\theta$ and $\rho$ are singlets. Note that in SYM the scalars transform in the $\rep{8}_v$ of $SO(8)$ which decomposes
%$$ \rep{8}_v \rightarrow \rep{6}_0 \oplus \rep{1}_{1/2} \oplus \rep{1}_{-1/2}$$
%under $SU(4)\times U(1)_b\subset SO(8)$. The scalars and fermions ABJM transform in the $\rep{4}\oplus %\brep{4}$ of $SU(4)$, which is not compatible with the representation theory above.

The rewriting of $\IR^8$ as $\IC^4$ is trivial with the two membrane locations denoted by $SU(4)$ vectors $\vec{z},\vec{w}$, and by translation invariance we identify the relative coordinate $\vec r=\vec{z}-\vec{w}$. Thanks to the classical $U(1)_b$ gauge symmetry the ABJM instanton calculation is smeared over the common phase circle of the $\IC^4$ coordinates of $\vec r$.  Geometrically, the $\IR^8$ of the relative coordinate is described as a cone over $S^7$, and the smearing averages over the circle fiber in $S^7$, written as a Hopf fibration with base $\IC\IP^3$.  The cone projected on this circle has a radial coordinate whose squared length is $m_{12}$.

To take into account that the moduli space is an orbifold $\IC^4/\IZ_k$, we sum over images on the original cone.  As the branes interact in a pairwise manner, by symmetry we can treat one of the membranes, say $\vec{z}$, as fixed and sum over the $k$ images of $\vec{w}$. Thus, the scalar prefactor in the vertex \eqref{vertex1} becomes
$$
\frac{1}{r^{10}} = \sum_{l=0}^{k-1} \frac{1}{|\vec{z}-e^{2\pi i l / k}\vec{w}|^5}.
$$
To relate this to the ABJM 1-instanton vertex, we Fourier transform in the $U(1)$ direction given by the phase separation of the two membranes,
$
e^{i\theta} = \frac{\vec{z}\cdot\vec{w}} {|\vec{z}||\vec{w}|}.
$
As the fermions are invariant under this $U(1)$, we only need to focus on the scalar prefactor whose Fourier modes are given by
$$
\sum_{l=0}^{k-1} \frac{1}{|\vec{z}-e^{2\pi i l / k}\vec{w}|^5} = \sum_p f_p(\vec{z},\vec{w}) e^{ikp\theta},
$$
with the inverse relation
\beq
f_p(z,w) = \sum_{l=0}^{k-1} \frac{k}{2\pi} \int_0^{2\pi/k} \frac{e^{-i p \theta} d\theta}{|\vec{z}-e^{2\pi i l / k}\vec{w}|^5}.
\eeq
To evaluate this integral we first note that $f_p=0$ unless $p\in k\IZ$. We then trade the sum over images for an extension of the domain of integration to $[0,2\pi]$. Finally  using $|\vec{z}-\vec{w}|^2 = |\vec{z}|^2 + |\vec{w}|^2 - 2 |\vec{z}\cdot\vec{w}^*|\cos\theta,$ the Fourier modes become
\beq
f_p(z,w) =  \frac{k}{2\pi} \int_0^{2\pi} \frac{e^{-i k p \theta} d\theta}{(|\vec{z}|^2 + |\vec{w}|^2 - 2 |\vec{z}\cdot\vec{w}^*|\cos\theta)^5}.\label{inverseFourier2}
\eeq
We solve this using $\cos \theta = (e^{i\theta} + e^{-i\theta})/2$ and changing variables to $q=e^{i\theta}$. Then  \eqref{inverseFourier2} becomes a contour integral, with the contour $C$ being the unit circle with the origin at zero:
\beq
f_p= \frac{k}{2\pi i} \oint_C \frac{q^{-pk-1}dq}{(|\vec{z}|^2 + |\vec{w}|^2 -  |\vec{z}\cdot\vec{w}^*|(q+q^{-1}))^5}.
\eeq
Let $\a = |\vec{z}|^2 + |\vec{w}|^2$, $\b=|\vec{z}\cdot \vec{w}^*|$. Then,
\beq
f_p= -\frac{k}{4!}\frac{d^4}{d\a^4} \oint_C \frac{dq}{2\pi i} \frac{q^{-k p}}{\b-\a q + \b q^2}.
\eeq
The countour integral has a contribution at $q=0$ and $q=q_-$, where
$$
q_\pm = \frac{\a}{2\b}\left(1 \pm \sqrt{1 - \frac{4\b^2}{\a^2}}\right).
$$
Using Cauchy integral formula and restricting to the case where the two membranes are in a single complex plane, we find for $p>0$:
\beq
\begin{split}
f_p &= \frac{k}{4!}\left(\frac{|z|}{|w|}\right)^{kp}\frac{1}{m^9} \Big[ (9-10k^2p^2 + k^4p^4)m^4 + (55kp-10k^3p^3)(|z|^2+|w|^2) m^3 +\cr \qquad\qquad &+ 45(k^2p^2-2)(|z|^2+|w|^2)^2m^2 -105(|z|^2+|w|^2)^3 kp m + 105(|z|^2+|w|^2)^4 \Big],
\end{split}\label{sugra1}
\eeq
where $m^2 = (|z|^2 - |w|^2)^2$, and we have assumed $|z|\le |w|$. The instanton result is a 1-loop semi-classical calculation, and to compare  \eqref{sugra1} with \eqref{correlatorfinal} we take the leading term in an expansion of $1/m$ is given by
\beq
f_p = \frac{k}{4!}\left(\frac{|z|}{|w|}\right)^{kp}\frac{k^4p^4-10k^2p^2 +9}{m^5} .
\eeq
The higher order terms in the $1/m$ expansion correspond to higher-loop corrections about the instanton. Finally, we extract the semiclassical charge 1 instanton result
\beq
f_1 = \frac{k}{4!}\left(\frac{|z|}{|w|}\right)^{k}\frac{k^4-10k^2 +9}{m^5} .
\eeq
Up to an overall normalization, the charge one vertex takes the form
\beq
\int d^3 x \left(\frac{|z|}{|w|}\right)^{k}\frac{1}{m^5} \big[a_1(\psi_1 \gamma^{pr} \psi_1)^2(\psi_2 \gamma^{qr} \psi_2)^2 + a_2(\psi_1 \gamma^{pr} \psi_1 \psi_2 \gamma^{pr} \psi_2)^2 \big],\label{vertex2}
\eeq
We are now in a position to compare with the result in \eqref{correlatorfinal}.  We first note that both results have the same power of $r$.  The correlator leading to \eqref{correlatorfinal} has four scalar insertions in addition to the eight fermions. The Wick contractions of the fermions will result in the scalars in the numerator, evaluated on their vev, combining to $m^2$; with the prefactor scaling as $m^{-5}$, the scaling of the instanton correlator \eqref{correlatorfinal} agrees with that of the vertex \eqref{vertex2}.   The $U(1)$ charges also agree, not surprisingly.

In the course of their analysis, \cite{Paban:1998mp} used an intricate series of Fierz identities to write the 8-fermion vertex in terms of the spinor bilinears in \eqref{vertex1} having explicit $SO(2,1)$ index structure. Consequently, \eqref{vertex1} is not manifestly $SO(2,1)$ Lorentz invariant, so to compare to \eqref{correlatorfinal} we need to delve into the index structure of the fermions.

Decomposing the contracted fermion bilinears $\psi\gamma^{pr}\psi$ under $SU(4)$, the only terms that can contribute when the vev is in a single complex plane have the structure
\beq\label{su4bilinears}
(\psi_{P\,\alpha}Z^P \bar\psi^Q_\beta) \times (\epsilon_{QRST} \bar\psi^R_\gamma Z^S \bar\psi^T_\delta ),
\eeq
where $Z^P = \phi^{2P-1}+i\phi^{2P}$. This agrees with the structure of \eqref{su4fermions} up to the specializations $\alpha=\beta$ and $\gamma=\delta$ in each of the two terms in \eqref{vertex2}. Note that $Z,\psi$ are coordinates on the moduli space, and hence have no matrix structure. 
The instanton measure is a product of two copies of the $SU(4)$ invariant structures \eqref{su4fermions}.  Specializing to a scalar vev restricted to a single complex plane, again only the first structure appears.  The $SU(4)$ indices on the $\bar\Psi$ fields are totally antisymmetric, therefore the $SO(2,1)$ indices are totally symmetric and comprise the spin-3/2 representation; the possible labels are $\beta\gamma\delta=111$, $112$, $122$ and $222$.
Using (anti)symmetry, one can readily arrange this specialization; the case where $\beta\gamma\delta=111$ corresponds to the $a_1$ term in \eqref{vertex2}, while the $\beta\gamma\delta=112$ case yields the $a_2$ term (with the index structures $222$ and $122$, respectively, are equivalent since they must appear in the other fermion quadrilinear in the product). 
The $112$ state has three terms relative to the $111$ state with an overall normalization of $1/\sqrt{3}$. Then, we find the ratio $a_1/a_2 = 1/3$. Thus we see complete agreement between the instanton calculation and the M-theory effective vertex, up to an overall normalization; and the determinantal structure of the instanton zero modes serves to fix the relative coefficient $a_1/a_2$ in the M-theory vertex.

Note that for this argument it is crucial that we have specialized to the relative coordinate on the moduli space, otherwise the fermions carry additional labels, voiding the symmetry structure.  It is also worth noting that the way in which the relative coordinate arises is more obscure in ABJM than in SYM.  ABJM is bi-fundamental $U(2)\times U(2)$ gauge theory while SYM is an adjoint $SU(2)$ gauge theory, and this difference is manifested in the IR limit in a number of ways. In SYM the separation of the overall U(1) is clean thanks to translation invariance and the adjoint nature of the fields: the centre of mass modes decouple from the modes describing the pairwise membrane interactions. The effective field theory, and corresponding vertex \eqref{vertex1}, can be written purely in terms of the light scalar $\phi$ describing the relative motion of the membranes. In ABJM, the $U(1)_b$ corresponding to the centre of mass modes does not easily decouple thanks to the bifundamental nature of the theory and the off-diagonal Chern-Simons couplings. The effective dynamics is then most straightforwardly described in terms of the two light scalars $\vec{z},\vec{w}$, which do not have a simple relation to $\phi$.

We do not yet have a complete calculation in the gauge theory showing the decoupling of the off-diagonal modes in the effective action, as that would require summing all of the monopole-instanton corrections. However, the supergravity analysis and $p=1$ monopole-instanton calculation does give an indication that the off-diagonal modes are a red herring and will decouple from the effective theory at low-energies when all the instanton effects are included.  The obvious next step is to perform the all-instanton calculation, just as \cite{Dorey:1997tr,Paban:1998mp} extended the one-instanton calculation of \cite{Polchinski:1997pz} in SYM. As discussed above, this could shed light on the type of interactions that would lift the $U(1)_b$ degeneracy of the light off-diagonal excitations.  Thanks to the $\Z_k$ quotient, the moduli space of $U(2)\times U(2)$ ABJM has three distinguished points: the two M2-branes and the orbifold point. This is in contrast with $SU(2)$ SYM whose moduli space has a single distinguished point, the centre of mass of the D2-branes. It would be interesting to explore analogies with monopole--instantons in say $SU(3)$ SYM \cite{Fraser:1997xi}, which has three distinguished points.

%%%%%%%%%%%%%%%%%%%%%%%%%%%%%%
%%%%%%%%%%%%%%%%%%%%%%%%%%%%%%

\section*{Acknowledgements}
We would like to thank S. Sethi, N. Dorey, S. Lee, A. Royston  and I. Melnikov for helpful discussions. EM is supported in part by DOE grant DE-FG02-90ER-40560. JM is supported by an EPSRC Postdoctoral Fellowship  EP/G051054/1.

\newpage
\appendix

%%%%%%%%%%%%%%%%%%%%%%%%%%%%%%
%%%%%%%%%%%%%%%%%%%%%%%%%%%%%%

\section{Massless Excitations in M2-brane theories}
\label{app:singular}
In this appendix we illustrate how the singular locus appears in other descriptions of M2-brane theories.

%%%%%%%%%%%%%%%%%%%%%%%%%%%%%%
%%%%%%%%%%%%%%%%%%%%%%%%%%%%%%

\subsection{Bagger-Lambert-Gustavsson}
\label{app:BLG}
At the level of the classical Lagrangian, it is straightforward to map the BLG theory to the $\SU(2)\times\SU(2)$ ABJM theory by a field redefinition \cite{Aharony:2008ug}. This was also observed \cite{Lambert:2010ji}  where subtleties with the Chern-Simons level $k$ and quantisation of flux were pointed out.

The BLG theory can be written as a $\SU(2)\times\SU(2)$ gauge theory with manifest $\N=8$ supersymmetry and $\SO(8)$ R-symmetry \cite{VanRaamsdonk:2008ft}. The scalars are denoted by $X^I$ for $I=1,\ldots,8$ and are related to the complex scalars of ABJM by
\be
Z^P = X^P + i X^{P+4}\label{eqn:transf_1}
\ee
for $P=1,2,3,4$. The scalars $X^I$ are in the bi-fundamental of $\SU(2)\times\SU(2)$ and obey a reality constraint $X^I = -\ve (X^I)^* \ve$ where $\ve = i\sigma^2$. The $X^I$ can then be parametrized as
\be
X^I = x^I_a \sigma^a, ~~ {\rm where}~~ \sigma^a = (1, i \vec{\sigma})
\ee
where $x^I_a$ are real numbers. We can define two operations that conjugate the $\SU(2)\times\SU(2)$ gauge symmetry representation and the $\SU(4)$ R-symmetry representation:
\bea
Z^{\dagger P} &=& - \ve (Z^P)^T \ve = X^{\dagger P} + i X^{\dagger P+4}, \quad {\rm gauge~symmetry}\cr
{Z}_P &=& - \ve (Z^P)^* \ve = X^P - i X^{P+4},\quad{\rm R-symmetry}.
\eea
These two operations can only be performed separately for a gauge group $\SU(2)\times \SU(2)$, where a reality constraint can be imposed. For $\U(N)\times \U(N)$ theories, only the combination $\Zb_P$ makes sense. We can invert
\bea
X^P  = \half ( Z^P + {Z}_P), \quad X^{P+4} = \frac{1}{2i} (Z^P - {Z}_P).\label{eqn:transf_2}
\eea
The potential in the BLG theory is
\be
V(X) = \frac{8}{3} \tr \left(X^{[I} X^{J \dagger} X^{K]} X^{K \dagger} X^{J} X^{I \dagger}\right)\ ,
\ee
and one can readily check that this potential, and indeed the entire BLG action, maps to the ABJM action under the field redefinition \eqref{eqn:transf_1}.
The BLG potential vanishes when the $X^I$ are diagonal:
\be
\vev{X^I} = x^I_0 + i x^I_3 \sigma^3.\label{eqn:VEV_1}
\ee
By an $\SO(8)$ rotation, we can put $x_0^I = x_3^I = 0$ for $I\ne1,5$. The two vectors $x_0^I$ and $x_3^I$ span the 1-5 plane in $\RR^8$. Under \eqref{eqn:transf_1} the 1-5 plane becomes the $\IC$-plane spanned by $\vev{Z^1}$. Expanding the potential to quadratic order:
\be
V(X) \sim \sum_{i,j=1}^4(x_0^1 x_3^5 - x_0^5 x_3^1)^2 (\delta X_{ij}^K)^2 + \ldots
\ee
with $K\ne 1,5$ and the mass of the lightest excitation goes like
\bea
m^2 &\sim& |x_0^1 x_3^5 - x_0^5 x_3^1|^2,\cr
&\sim& |\vec{x}_0 \times \vec{x}_3|^2.\label{trianglemass1}
\eea
which is the area of the triangle spanned by $\vec{x}_0$, $\vec{x}_3$ and the origin in 1-5 plane. In particular note there are massless scalar excitations when $\vec{x}_0$ and $\vec{x}_3$ become collinear. How does this compare with \eqref{m2mass_2}? Apply the transformation \eqref{eqn:transf_1} to \eqref{eqn:VEV_1}:
\be
\vev{Z^1} = \begin{pmatrix} x^1_0 - x^5_3 + i (x^1_3 + x^5_0) & 0 \\
0 & x^1_0 + x^5_3 - i (x^5_0 - x^1_3)\end{pmatrix}
\ee
We identify
\be
z^1 =  x^1_0 - x^5_3 + i (x^1_3 + x^5_0), ~~{\rm and}~~ w^1 = x^1_0 + x^5_3 - i (x^5_0 - x^1_3).\label{eqn:cpx_1}
 \ee
With this identification, the two formulae \eqref{m2mass_2} and \eqref{trianglemass1} agree. However, the interpretation of the coordinates of the moduli spaces in the ABJM and BLG theories are different. In the ABJM theory the VEVs $z^P$ and $w^P$ are interpreted as the coordinates of the two M2-branes. In \cite{Lambert:2008et,VanRaamsdonk:2008ft}, the VEVs $\vec{x}_0$ and $\vec{x}_3$ are interpreted as the coordinates of the two M2-branes. This change in interpretation together with \eqref{eqn:transf_1} maps a mass going like the area of a triangle \eqref{trianglemass1} into a mass going like the separation in radius \eqref{m2mass_2}.

%%%%%%%%%%%%%%%%%%%%%%%%%%%%%%
%%%%%%%%%%%%%%%%%%%%%%%%%%%%%%

\subsection{M2-branes probing toric $CY_4$-folds}
\label{app:cy4}
Although we explicitly analysed this behavior for the $\U(2)\times\U(2)$ ABJM theory, it persists for more general constructions. Firstly, it is clear the analysis above generalises to $\U(N)\times\U(N)$ theories. Secondly, the theories corresponding to M2-branes probing non-compact Calabi-Yau's also have this behavior. These theories are constructed using a generalization of the tiling techniques familiar from Hanany-Witten constructions in 3+1 dimensions. We will now review the pertinent features of these constructions and refer the reader to any of the original references for more details, for example \cite{Hanany:2008cd}.

The tiling constructions give rise to  2+1-dimensional theories with $\N=2$ supersymmetry, Chern-Simons couplings and a product of $\U(N)$ gauge groups $G_1\times\ldots\times G_r$. The Chern-Simons terms appear in such a way that $\sum_{i=1}^r k_i =0$. The theory flows in the IR to a non-trivial fixed point; the moduli space is given by solving the D-terms and F-terms modulo gauge transformations. The theory is then argued to describe M2-branes probing a non-compact Calabi-Yau cone.

Our particular interest is in the moduli space. The bosonic potential is given by
 \be
 V \sim \Tr \left[-4 \sum_i k_i \sigma_i D_i + \sum_i D_i \mu_i(Z) - \sum_{Z_{ij}} \left|\sigma_i Z_{ij} - Z_{ij}\sigma_j\right|^2 - \sum_{Z_{ij}} \Big| \del_{Z_{ij}}W\Big|^2\right]
 \ee
where $Z_{ij}$ is a scalar field in the bifundamental of $G_i\times G_j$; $\sigma_i$ is an auxiliary field in the corresponding 2+1 vector supermultiplet; and $\mu_i(Z)$ is the moment map action for the $a$-th gauge group given by
\bea
\mu_i(Z) = \sum_j \left(Z_{ij} \Zb_{ij}  -   \Zb_{ji} Z_{ji}\right)  + [Z_{ii} ,\Zb_{ii}].
\eea
The potential is a sum of squares and vacua are given by setting the last two terms to zero, and integrating out the D-terms, $D_i$. The vacuum conditions are then given by
\bea
\sigma_i Z_{ij} - Z_{ij}\sigma_j &=& 0,\cr
\mu_i(Z) &=& 4k_i \sigma_i,\cr
 \del_{Z_{ij}}W &=& 0.\label{eqn:toric_vacua}
\eea
In the particular case where the gauge groups are all abelian, these equations are straightforward to solve. The first implies all the $\sigma_i$ are equal viz.  $\sigma_i = \sigma$ for $i=1,\ldots,r$. The second equation imposes the symplectic quotient of the toric variety. The last equation is the standard F-term constraint from 3+1 dimensions.

We now make a change of coordinates that illustrates the appearance of the peculiar loci we discussed above for the $\U(2)\times\U(2)$ ABJM theory. The change of coordinates is defined by the $r\times r$ matrix
\be
M_{ij} = \begin{pmatrix}m_1 & m_2 & \ldots & m_r\\
1 & 1 &\ldots &1\\
M_{31} & M_{32} & \ldots & M_{3r}\\
&\ldots&\ldots
\end{pmatrix}\label{m}
\ee
This matrix has the property that all the rows are orthogonal and $\vec{k} = \lambda\vec{m}$ for some $\lambda\in\Z_{>0}$.  The new gauge fields are defined by $\wt{A}_{(i)} =  \sum_{j=1}^r M_{ij}A_{(j)} $, and denote $\wt{G}_j$ the gauge group corresponding to $\wt A_{(i)}$.%
\footnote{To recover the ABJM analysis in the previous subsection, we set $r=2$ and $\vec{m}=(1,-1)$. Then $\wt G_1 = \U(1)_b$ and $\wt G_2 = \U(1)_D$.} The Chern-Simons term is
\bea
S_{CS} &=& \sum_i \frac{k_i}{4\pi} \int_\Sigma A_{(i)} \w dA_{(i)},\cr
&=& \sum_{i,j,k} \frac{k_i}{4\pi} M^{-1}_{ij} M^{-1}_{ik} \int_\Sigma \wt{A}_{(j)} \w d\wt{A}_{(k)},\cr
&=& \frac{\lambda}{2\pi r }\int_\Sigma \wt{A}_{(1)} \w d\wt{A}_{(2)} + \sum_{i,j\ne 2,k\ne 2} M^{-1}_{ij} M^{-1}_{ik} \frac{k_i}{4\pi}\int_\Sigma \wt{A}_{(j)} \w d\wt{A}_{(k)}.\label{hananyCS}
\eea
In the last line we used the orthogonality of the rows of $M_{ij}$.
The covariant derivative becomes
\bea
D Z_{ij} &=& dZ_{ij} - iA_{(i)} Z_{ij} + i Z_{ij} A_{(j)},\quad({\rm no~sum~on~i,j}),\cr
 &=& dZ_{ij} - i\sum_{k\ne2} (M^{-1}_{ik} - M^{-1}_{jk}) \wt{A}_{(k)} Z_{ij},
\eea
where $\wt{A}_{(2)}$ dropped out due to the structure of \eqref{m}, and consequently the gauge group $\wt{G}_2$ is unbroken in the vacuum. The remaining $r-1$ gauge groups $\wt{G}_1$ and $\wt{G}_i$ with $i\ge 3$ are Higgsed for generic expectation values for the scalars $Z_{ij}$.

Using $\sum_i k_i = 0$ and $\sum_i \mu_i = 0$ we see one of the D-term constraints is redundant. The remaining $r-1$ D-term constraints become
\bea
4 (\vec{m}\cdot\vec{k}) \sigma &=& \vec{m}\cdot\vec{\mu}(Z), \label{eqn:sigma}\\
\wt{\mu}_i(Z) &=& M_{ij} \mu_j(Z) = 0,\qquad{\rm ~for~}i\ge 3.\label{eqn:mom}
\eea
The first equation determines the field $\sigma$ in terms of the scalars $Z_{ij}$ and does not constrain the moduli space. The remaining $r-2$ equations impose constraints on the fields $Z_{ij}$ via the moment map. As happens in 3+1 dimensions the $r-2$ D-term constraints and the action of the $r-2$ gauge groups $\wt G_{i}$ for $i\ge 3$ may be imposed as a complexified gauge quotient. The moduli space is then a toric Calabi-Yau four-fold realised as a holomorphic quotient.

 Thus, in these generalizations of ABJM, there are two distinguished gauge groups $\wt G_1$ and $\wt G_2$ and play a role analogous to the $\U(1)_b$ and $\U(1)_D$ respectively in ABJM.

The analysis of the non-abelian moduli space follows from the abelian analysis in a similar manner to the $\U(N)\times\U(N)$ ABJM theory [cf. the discussion in section (\ref{sect:modulispace})]. Indeed, by studying equations in \eqref{eqn:toric_vacua} (in a gauge where the $\sigma_i$ are diagonal), a generic solution has the form of $Z_{ij}$ being diagonal%
\footnote{Diagonal $Z_{ij}$ is certainly a sufficient condition for $V=0$. There remains the possibility however of more general solutions to the vacuum equations in which the $Z_{ij}$ are not diagonal \cite{Hanany:2008cd} in which case the moduli space dynamics may be more interesting.}. The scalar fields then Higgs the gauge groups down to a symmetric product of the abelian groups. This is the generalization of the $\U(N)\times\U(N)$ ABJM theory breaking to $\U(1)^N \times \U(1)^N$ via the expectation values in \eqref{VEV_2} .

The non-abelian generalization has one important feature: the theory can now carry monopole--instantons. To see these we write out the schematic equations of motion for the gauge fields. They are given by
\bea
 \star \wt F_{(1)}  &=& 0, \cr
\frac{\lambda}{2\pi r} \star \wt F_{(2)}  &=& \wt j_{(1)} -\sum_{i;k\ne 2} M^{-1}_{i1} M^{-1}_{ik} \frac{k_i}{4\pi}  \wt F_{(k)}, \cr
\sum_{i,j\ne 2}  M^{-1}_{ij} M^{-1}_{ik} \frac{k_i}{4\pi}  \wt F_{(j)}&=& \wt j_{(k)}, \quad {\rm for~}k>2\label{CSEOM4}
\eea
with the matter currents defined as
\be
\wt j_{(k)} = -i \sum_{i,j}\sum_{k\ne 2}(M^{-1}_{ik} - M^{-1}_{jk}) \big(Z_{ij} D \Zb^{ij} - \Zb^{ij}DZ_{ij} \big)
\ee
Again, we see the distinguished role of $\wt G_{1}$ and $\wt G_{2}$, as is to be expected from the structure of the D-terms and \eqref{m}.

What does \eqref{CSEOM4} tell us about the structure of monopoles in these theories? Suppose we are in a monopole background. Then, as all of the gauge groups except $\wt{G}_{2}$ are Higgsed, the monopole can only have its field strength in $\wt G_2$. Hence, $\wt F_{j}=0$ for $j\ne 2$. In this case, \eqref{CSEOM4} reduces to the same set of equations as for the ABJM theory. Thus, we expect monopole--instantons to play an identical role in the generalisations of ABJM to the original construction. Also,  as $\wt{A}_{(2)}$ only appears in the Lagrangian via its field strength $d\wt{A}_{(2)}$ in the Chern-Simons term \eqref{hananyCS} it may be dualized to a scalar in the same way as the $\U(1)_D$ field strength in ABJM. Monopole--instantons then imply $\wt{G}_1$ is broken to a discrete subgroup $\Z_{\lambda} = \Z_{{\rm gcd}(k_1,\ldots,k_r)}$. Finally, $\wt{G}_1$ is a symmetry of the classical Lagrangian. Hence, by the same argument we gave above for ABJM, this implies there is a $\U(1)$ locus in the moduli space along which there are anomalous massless excitations. Thus, although we have focussed in this note on the ABJM $\U(2)\times\U(2)$ theory, it is clear our analysis generalises to more involved theories.

%%%%%%%%%%%%%%%%%%%%%%%%%%%%%%
%%%%%%%%%%%%%%%%%%%%%%%%%%%%%%

\section{Monopole-Instanton Fluctuation Determinant}
\label{app:1loop}
This section outlines the computation of the small-fluctuation determinant about the monopole-instanton background.

%%%%%%%%%%%%%%%%%%%%%%%%%%%%%%
%%%%%%%%%%%%%%%%%%%%%%%%%%%%%%

\subsection{Bosonic Fields}
The bosons, $Z^P,A_{(1)},A_{(2)}$, are involved in three contributions: the scalar kinetic terms, bosonic potential and Chern-Simons term. We write the gauge bosons in terms of their diagonal and baryonic linear combinations: $A_D = \half (A_{(1)} + A_{(2)}), A_b = \half (A_{(1)} - A_{(2)})$ and consider each of the pieces in turn.

%%%%%%%%%%%%%%%%%%%%%%%%%%%%%%
%%%%%%%%%%%%%%%%%%%%%%%%%%%%%%

\subsubsection*{Scalar kinetic Terms}
These are of the form
$$
- \tr D_\mu \Zb_P D_\mu Z^P.
$$
Expand to quadratic order, keeping in mind that we need to consider both $\dA$ and $\dAt$ fluctuations. To quadratic order we find
\beq
\begin{split}
D_\mu Z^P ~=~& \hat D_\mu \delta Z^P-2ia\delta^{P1} \dAt_\mu  - i[\delta A_\mu , \delta Z^P] -  i\{\dAt_\mu , \delta Z^P\}, \cr
D_\mu \Zb_P ~=~& \delta_{P1} b \Big(\hat D_\mu \Phi - i [\delta A_\mu , \Phi] -  i\{\dAt_\mu,\Phi\}\Big)  -2ic\delta_{P1} \dAt  - i [ \dA_\mu,\delta \Zb_P ]  +\cr
& \qquad \qquad - i\{\dAt_\mu,\delta \Zb_P \} + \hat D_\mu \delta \Zb_P
\end{split}
\eeq
where where $\hat D$ is the connection computed with respect to the monopole-instanton background where $A_{(1)}=A_{(2)}=A_D$.  Plugging these expressions into the kinetic term and keeping up to quadratic pieces we find
\be
\begin{split}
-\tr D^\mu Z^P D_\mu \Zb_P ~=~& \tr\Big(\,  \delta Z^P \hat D^\mu\hat D_\mu \delta \Zb_P  +2ib [\delta A_\mu, \delta Z^1] \hat D_\mu \Phi -2ab \dAt_\mu [\Phi, \dA_\mu]\cr
&    + 4a\dAt_\mu (b\Phi+cI_2) \dAt_\mu + ib \dZ^1 [ \Phi,F]\Big).\label{ke}
\end{split}
\ee
In deriving this we've integrated by parts and made use of the gauge fixing condition
\be
\hat D_\mu \dA_\mu = F(\dZ,\dZb,\Phi),\quad \hat D_\mu \dAt_\mu = 0.\label{gaugefixing1}
\ee
We need to find a convenient choice for $F$. It turns out that if $F = 0$ or $F\propto [\Phi,\delta Z]$, the determinant is unchanged. However, the zero-mode analysis easiest in the gauge $F= 2 i mb^{-1}\Phi \delta \Zb$.

%%%%%%%%%%%%%%%%%%%%%%%%%%%%%%
%%%%%%%%%%%%%%%%%%%%%%%%%%%%%%

\subsubsection*{Scalar Potential}
In addition to its expression in \eqref{pot}, the potential can also be written as a sum of a D-term and an F-term \cite{Benna:2008zy}:
\begin{equation}
\begin{split}
V_F =& -\frac{16\pi^2}{9k^2} \epsilon_{PQRV}\epsilon^{STUV} \tr Z^{\ddag\,P} Z^Q Z^{\ddag,R} \Zb_S \Zb^{\ddag}_T \Zb_U, \cr
V_S =& \frac{4\pi^2}{k^2} \tr \Big[Z^P \Zb_P Z^Q \Zb_Q Z^R \Zb_R + \Zb_P Z^P \Zb_Q Z^Q \Zb_R Z^R -2 \Zb_P Z^Q \Zb_Q Z^P \Zb_R Z^R \Big].
\end{split}
\nonumber
\end{equation}
To quadratic order, the F-term does not contribute while the D-term gives
\bea
\delta V_D &=& -\frac{4\pi^2 a^2b^2}{k^2} \tr [\delta Z^{P'}, \Phi] [\delta \Zb_{P'} , \Phi],\label{scalar2}
\eea
$P'=2,3,4$. Note the $\dZ^1$ fluctuations cancelled out without the need to impose any gauge conditions or choices of polarisation.

In analogy to the $d=4$  SYM monopole \cite{Dorey:1997ij} we can formally rewrite this in terms of four-dimensional quantities. Introduce a four-dimensional index $m=1,2,3,4$ and define  a four-dimensional operator
\be\label{curlyD}
\cD_m = (\hat D_\mu,  D_4), \quad {\rm where~} \cD_4 \delta Z^{P'} = -i[A_4,\delta Z^{P'}].
\ee
We can then expression the fluctuation of $V_D$ as
\be
\delta V_D = - \tr \delta \Zb_{P'} \cD_4^2 \delta Z^{P'}
\ee

%%%%%%%%%%%%%%%%%%%%%%%%%%%%%%
%%%%%%%%%%%%%%%%%%%%%%%%%%%%%%

\subsection*{Chern-Simons and gauge fixing}
The Chern-Simons Lagrangian is given by \eqref{csaction}, and we expand it to quadratic order:
\bea
\cL_{CS} &=& \frac{ik}{4\pi} \tr\Big[ (\delta A_1 F^{(1)}_2 - \delta A_2 F^{(2)}_2) + (\delta A_1 \hat D^{(1)} \delta A_1 - \delta A_2 \hat D^{(2)} \delta A_2)\Big],
\eea
where $\hat D^{(i)} \delta A_i = d \delta A_i - i [A_{(i)} ,\delta A_i]$. To clarify, background gauge bosons are denoted $A_{(1)},A_{(2)}$ while fluctuations are denoted $\delta A_1, \delta A_2$. The linear terms cancel once we apply the equations of motion for the scalars, and so we will not worry about them. Rewriting in terms of the diagonal and baryonic basis we get
\bea
\cL_{CS} &=& \frac{ik}{\pi} \tr\Big[  \delta A_D\wedge \hat D \delta  A_b\Big]
\eea
where $\hat D  \dAt = d \dAt - i A \wedge \dAt$.

There is also a gauge fixing condition which we can impose using a Lagrange multiplier:
\be
S_{\rm gauge~fixing} = \int d^3 x \lambda_1  D_\mu A_{D\,\mu}  +\lambda_2 D_\mu A_{b\,\mu}
\ee
The fields $\lambda_1,\lambda_2$ will also contribute to the scalar determinants.

%%%%%%%%%%%%%%%%%%%%%%%%%%%%%%
%%%%%%%%%%%%%%%%%%%%%%%%%%%%%%

\subsubsection*{Scalar determinant}
We now put the pieces together to compute the determinant. It naturally splits into studying fluctuations orthogonal to the plane of the VEV and those along the plane of the VEV. The fluctuations orthogonal to the plane of the VEV are of the form $\tr \dZb_{P'} \cD_m^2 \dZ^{P'}$ and give rise to a determinant of the Laplacian built from the derivative operator~\eqref{curlyD} in the monopole background
\be
(\det \cD^2_m)^{-3}\label{scalardet1}
\ee
The fluctuations in the plane of the VEV $\delta Z^1$ are grouped together with the gauge bosons, and $\lambda_1,\lambda_2$. With our choice of gauge, the term proportional to $F$ in \C{ke} is
$
\tr ib \dZ^1 [ \Phi,F] = -\tr \delta \Zb_1 \mathcal{D}_4^2 \delta Z^1.
$
The quadratic fluctuation operator may then be written as a bilinear form
$$
-\tr D_\mu Z^1 D_\mu \Zb_1 + \cL_{CS} + \cL_{\rm gauge fixing} = \Big(\dA_\mu ~\dAt_\nu ~\dZ^1 ~\dZb_1 ~\lambda_1~\lambda_2\Big) ~\Delta^{\mu\nu;\rho\tau} ~\begin{pmatrix}\dA_\rho\\ \dAt_\tau\\ \dZ^1 \\\dZb_1\\ \lambda_1\\ \lambda_2 \end{pmatrix}
$$
where the operator $\Delta$ takes the form
\bea
\Delta^{\mu\nu;\rho\tau} = \begin{pmatrix}
0 & 2ab \Phi \delta^{\mu\tau} + \frac{k}{2\pi} \e^{\mu\lambda\tau}D_\lambda & -2ib D_\mu \Phi & 0 & -D_\mu & 0\\
-2ab \Phi \delta^{\nu\rho} + \frac{k}{2\pi} \e^{\nu\lambda\rho}D_\lambda & 4a(b\Phi+cI_2) \delta^{\tau\nu} & 0& 0 &0 &-D_\nu \\
2ib D_\rho \Phi & 0 & 0 & \half \mathcal{D}_m^2 & 0 &0 \\
0 & 0 & \half \mathcal{D}_m^2 & 0&-2i\frac{m^2}{b}\Phi&0\\
D_\rho & 0 & 0  & 2i\frac{m^2}{b}\Phi& 0 & 0 \\
0&D_\tau & 0 & 0 & 0 &0  \\
\end{pmatrix}
\nonumber
\eea
The determinant is remarkably nice: $\det \Delta = (\det \cD_m^2)^2 (\det D_\mu^2)^4$, and together with \C{scalardet1}, we find the 1-loop determinant of the scalars is
\bea
\delta Z^P+\delta \Zb_P}+{\rm gauge~bosons+\lambda_{1,2} &=& (\det D_\mu^2)^{-2} (\det \cD_m^2)^{-4}
\label{scalardet2}
\eea
where we have inverted and taken the appropriate square root.

%%%%%%%%%%%%%%%%%%%%%%%%%%%%%%
%%%%%%%%%%%%%%%%%%%%%%%%%%%%%%

\subsection{Fermionic Fields}
The fermions have vanishing expectation value, and their kinetic term is
\be
 i\tr(\ol{\dpsi}^{P}\gamma^\mu \hat D_\mu \dpsi_P).
\ee
The only non-trivial Yukawa type term is given by
\bea
\delta V_D &=& \frac{2\pi iab}{k} \tr \Big[\dpsib_{P'} [ \Phi, \dpsi^{P'} ] - \dpsib_1 [\Phi,\dpsi^1]\Big]
\eea
As for the scalars, we introduce a formal gauge boson
\be
A_4 = \frac{2\pi ab}{k}\Phi,
\ee
and the covariant derivative is defined as
$$
\cD_m = (\hat D_\mu,  D_4), \quad {\rm where~} D_4 \dpsi = -i[A_4,\dpsi].
$$
The Weyl basis for the gamma matrices in positive signature flat metric for $\SO(4)$ consists of
$$\sigma^m = (\sigma^\mu, -i),\quad \bar \sigma^m = ( \sigma^\mu,i)$$
and the corresponding Dirac operators $\cDslash = \sigma^m \cD_m$ and $\cDslashb = \bar\sigma^m \cD_m$ . Then, the fermionic terms become
\bea
\delta \cL_{\rm ferm} &=&  \frac{i}{2}\tr\Big(\dpsib^{P'}\cDslash \dpsi_{P'} +  \dpsi_{P'}\cDslashb \dpsib^{P'}\Big) + \frac{i}{2}\tr \Big( \dpsib^{1}\cDslashb \dpsi_{1}+  \dpsi_{1}\,\cDslash \dpsib^{1}\Big)\label{fermions}
 \eea
In order to evaluate the Gaussian integration it is convenient to rewrite this as a 4D matrix
 \bea
\delta \cL_{\rm ferm}
&=& \frac{i}{2}\tr\Big(\dPsib^{P'}\begin{pmatrix} 0 & \cDslashb \\ \cDslash & 0\end{pmatrix}\dPsi_{P'} +  \dPsib^{1}\begin{pmatrix} 0 & \cDslash \\ \cDslashb & 0\end{pmatrix} \dPsi_{1}\Big) \label{fermions2}
\eea
where $\dPsib = \dPsib \Gamma^0$ is the usual Dirac conjugate. Letting
\be
\Delta_F =  \begin{pmatrix} 0 & \cDslashb \\ \cDslash & 0\end{pmatrix}, \quad \widetilde \Delta_F =  \begin{pmatrix} 0 & \cDslash \\ \cDslashb & 0\end{pmatrix}
\ee
the 1-loop determinant is then given by
\bea
(\det \Delta_F^2)^{3/4}(\det \wt\Delta_F^2)^{1/4} &=&  (\det \cDslash \cDslashb)(\det \cDslashb \cDslash)
\eea
The four-dimensional field strength $F_{mn}$ is defined to be the three-dimensional field strength together with $F_{\mu 4} = \hat D_\mu A_4$. As $\eps^{\mu\nu}_{~~~\rho}F_{\mu\nu} = \hat D_\rho A_4$, this field strength is self-dual $F_{mn} = \e_{mn}^{~~~pq}F_{pq}$. The operators above become
\bea
\cDslash \cDslashb &=& \cD_m^2 + \sigma^{mn} F_{mn}, \quad \sigma^{mn} = \frac{1}{4}( \sigma^m \sigb^n -  \sig^n \sigb^m),\cr
&=& \cD_m^2,\cr
\cDslashb \cDslash &=& \cD_\mu^2 + \sigb^{mn} F_{mn}, \quad \sigb^{mn} = \frac{1}{4}( \sigb^m \sig^n -  \sigb^n \sig^m),\cr
&=& \cD_m^2 + 2i\sig^{\mu} B_{\mu}.\label{dirac1}
\eea
We have used $\sigb^{mn}F_{mn} = i \sig^\mu B_\mu$, with $B_{\mu}$ the monopole field strength, and $\sig^{mn}F_{mn} = 0$. The fermionic determinants then become
\bea
{\rm 1-loop~Fermions}&=&  (\det I_{2\times 2}\cD^2 )(\det \cD^2 + 2i \sig^\mu B_\mu )\label{fermiondet1}
\eea
where $I_{2\times 2}$ is the rank 2 identity matrix in inserted explicitly to emphasise these terms are determinants of 2-component spinor operators.

%%%%%%%%%%%%%%%%%%%%%%%%%%%%%%
%%%%%%%%%%%%%%%%%%%%%%%%%%%%%%

\subsection{Ghosts}
The FP ghosts from the background gauge fixing give rise to
\be
{\rm FP} = (\det D^2)^2.\label{FPdet1}
\ee
Note this will nicely cancel the corresponding term from the gauge fixing in \C{scalardet2}.

%%%%%%%%%%%%%%%%%%%%%%%%%%%%%%
%%%%%%%%%%%%%%%%%%%%%%%%%%%%%%

\subsection{Final Result}

Putting together \C{scalardet2}, \C{fermiondet1}, and \C{FPdet1}, we get
\be
\begin{split}
\Dloop ~=~& [(\det \cD^2)^{-3-1} (\det D^2)^{-2}]\times[(\det I_{2\times2}\cD^2)(\det \cD^2 + 2i \sig^\mu B_\mu)]\times(\det D^2)^2 \cr
~=~& \frac{(\det \cD^2 + 2i \sig^\mu B_\mu)}{(\det I_{2\times 2}\cD^2)^2}.
\end{split}
\ee
In the first line, the first term is the bosonic determinant, the second term the fermionic determinant and the last term from the FP term. We can interpret this as the following. The field content of our theory is roughly the dimensional reduction of two $d=4$ $N=2$ hypermultiplets. Each hypermultiplet gives a factor of $$\Dloop^{-1}=\frac{(\det \cD^2 + 2i \sig^\mu B_\mu)^{1/2}}{(\det \cD^2)}.$$ The gauge multiplet completely cancels against the FP gauge fixing determinant, consistent with the fact that in Chern-Simons theories the gauge field has no propagating degrees of freedom. To contrast with 3D N=8 Yang-Mills the field content consists of a $d=4$ $N=2$ vector multiplet and a $N=2$ hypermultiplet reduced to 3D. The vector multiplet has propagating degrees of freedom and thus does not completely cancel against the gauge fixing determinant. Instead it gives a factor of $R$ which cancels against the $R^{-1}$ coming from the hypermultiplet, the end result is unity \cite{Dorey:1997tr}.

This determinant was evaluated in \cite{Dorey:1997ij}, which in our conventions and normalization is dimensionless, taking the form
\be
\Dloop ~=~ {2}^{-4}
\ee

%%%%%%%%%%%%%%%%%%%%%%%%%%%%%%
%%%%%%%%%%%%%%%%%%%%%%%%%%%%%%

\section{Zero Mode Integral}
\label{app:zeromode}
After integrating out the non-zero modes as is done in Appendix \ref{app:1loop}, the path integral reduces to a finite integral over the bosonic and fermionic zero modes of the background.

%%%%%%%%%%%%%%%%%%%%%%%%%%%%%%
%%%%%%%%%%%%%%%%%%%%%%%%%%%%%%

\subsection{Bosonic Zero Modes}
With the ansatz $A_L=A_R$, the only classical solutions are equivalent to the Yang-Mills BPS monopole monopole. Consequently, there are zero modes associated with the monopole solution. The gauge group is
\bea
\U(2)_D \times \U(2)_b = \U(1)_D \times \SU(2)_D \times \U(1)_b \times \SU(2)_b
\eea
The monopole solution spontaneously breaks the $\U(2)_b$ and the $\SU(2)_D\rightarrow \wt \U(1)_D$, where the $\widetilde \U(1)_D$ is the unbroken gauge group carrying the magnetic field strength. There are non-normalizable zero modes associated with the broken $\U(2)_b$ and the non-abelian W-bosons in $\SU(2)_D$, as well as motion of the vev in the $\RR^8$ moduli space. As usual these do not contribute to the monopole moduli space. For  monopole-instanton charge $p=1$, the monopole zero modes are three from translation of the centre of the monopole as well as a global $U(1)_D$ rotation. These are to be converted to an integral over the collective coordinates of the monopole, which induces a Jacobian. Schematically denote our fields as $\phi_a$ and moduli $m_i$. Then, the metric on field space $g_{ab}$ naturally induces a metric $h_{ij}$ on the moduli space
\be
h_{ij} = \int d^3 x \frac{\delta \phi_a}{\delta m^i} \frac{\delta\phi_b}{\delta m^j} g^{ab}.\label{modulimetric1}
\ee
The metric on field space we identify from kinetic terms in the Lagrangian, which for us is always a constant.  The integral over the zero modes then pulls backs to an integral over the moduli space as (see for example, \cite{Bernard:1979qt} for a more detailed discussion)
\be
\int \prod_i d\phi_i = \int \prod_r dm_r\,J(m), ~~ J(m) = \sqrt{\det h}
\ee
where $J(m)$ is a Jacobian from the change of coordinates.
For us, the Jacobian is straightforward to compute thanks to its close relation to the Yang-Mills BPS monopole. The metric on field space is diagonal, with the scalar field metric being $g_{P \ol{Q}} = \half\delta_{P\ol{Q}}$, while the gauge field metric is  $g_{A_L A_L} = g_{A_R A_R} = \frac{k}{4\pi}, g_{A_L A_R} = 0$. The three translation zero modes excite only $A_L,A_R,\Zb_1$ which can be seen by translating the monopole background \C{eqn:bog_1}, \C{monopole_1} by a constant vector $x^\mu \rightarrow x^\mu + v^\mu$, $v/m \ll 1$:
\bea
\delta A_{L\,\mu} &=& v^\nu \del_\nu A_L + D_\mu \Lambda\cr
 &=& v^\nu F^{\rm cl}_{\mu\nu}, \cr
\delta A_{R\,\mu}  &=& v^\nu F^{\rm cl}_{\mu\nu} \cr
\delta \Zb_1 &=& \del_\nu \Zb_1 - i [\Lambda, \Zb_1]\cr
& =& b  D_\nu \Phi.\label{eqn:zeromode1}
\eea
We have gauge transformed the fluctuations using $\Lambda = - v^\nu A_\nu$, and used $\Lambda \ll 1$ to drop the terms $[\delta A_\mu, \Lambda]$ and $-i[\Lambda,\delta \Zb]$ in the gauge transformation of the fluctuation. Note that $D_\mu \delta A_{L\,\mu} =  v^\nu D_\mu F^{\rm cl}_{\mu\nu} =  2 i m b^{-1} \delta \Zb \Phi$, which is compatible with the gauge choice \C{gaugefixing1} in the 1-loop calculation. We can now compute the metric on the moduli space parametrized by $v^{\mu}$
 \bea
 h_{ij} &=&  \frac{1}{V}\int d^3 x \big [ \frac{ik}{4\pi}\frac{\delta A_{L\,\mu}}{\delta v^i} \frac{\delta A_{L\,\mu}}{\delta v^j} +  \frac{ik}{4\pi}\frac{\delta A_{R\,\mu}}{\delta v^i}\frac{\delta A_{R\,\mu}}{\delta v^j} + 2g^{z \zbar} \frac{D \delta Z^1}{\delta v^i} \frac{D \delta \Zb_1}{\delta v^j} \big],\cr
 &=&\delta_{ij}\frac{ik}{4\pi V} \int d^3 x \, (F^{cl}_{\mu\nu})^2,\cr
 &=&  \delta_{ij} \frac{k m p}{V},
 \eea
 where in the last line $p$ is the monopole charge, $V$ is the volume of spacetime and we used the equation of motion $F_{\mu\nu} =-i m \epsilon_{\mu\nu\rho}D_\rho \Phi$, to convert the volume integral into an integral over the asymptotic two-sphere.

There is an additional zero mode global $\U(1)_D$ rotations. The monopole spontaneously breaks this symmetry, thereby generating a zero mode. This is most easily computed in singular gauge. In that case $Z \propto a I_2,$ and $\Zb = f(x) m \sigma_3$. Consequently, the $\U(1)_D$ gauge transformations commute with the scalar. The gauge fields on the other hand generate a zero mode
\bea
\delta A_L &=& \delta \theta\Big[\frac{i}{2}[\sigma_3,A_{L \, sing}]  + D (\Phi_{\rm sing} - \half \sigma_3)\Big],\cr
&=& \delta \theta D \Phi_{\rm sing}, \cr
\delta A_R &=& \delta \theta D \Phi_{\rm sing}.
\eea
where $\delta \theta$ is the infinitesimal $\U(1)_D$ parameter. The induced metric is similar to that computed above:
\bea
 h_{\theta \theta} &=&  \frac{1}{V}\int d^3 x \big [ \frac{i k}{4\pi}\left(\frac{\delta A_{L\,\mu}}{\delta \theta} \right)^2 +  \frac{i k}{4\pi}\left(\frac{\delta A_{R\,\mu}}{\delta \theta} \right)^2 + 2g^{z \zbar} \frac{D \delta Z^1}{\delta \theta} \frac{D \delta \Zb_1}{\delta \theta} \big],\cr
 &=& \frac{k}{4\pi V} \int d^3 x \, (D_\mu \Phi)^2,\cr
 &=&  \frac{kp}{V m}.
\eea
and $h_{\theta i} = 0$. Hence, the bosonic Jacobian for $p=1$ is
\bea
J_B &=& \frac{1}{V^2}  (k m)^{3/2}  \left(\frac{k}{m}\right)^{1/2},\cr
& =& \frac{k^2 m}{V^2}
\eea

The bosonic measure is normalized by demanding the Gaussian integral over the non-zero modes satisfies
\be
N_B \int \mathcal{D} \delta A \exp\left(-m\int d^3 x \delta A_\mu  \delta A_\mu \right) = 1.
\ee
This fixes the constant $N_B$ and this normalization then descends to the zero-mode integrals.  The power of $m$ in the exponential is fixed by demanding the exponent to be dimensionless. Putting all this together, the zero mode measure is
\be
\begin{split}
d\mu_B ~=~& \left( mV\pi\right)^{2}  \prod_{n=1}^4 \mathcal{D} \delta A_n,\cr
~=~&k^2 m^3 \pi^2 d^3 X_{cm} d\theta
\end{split}
\label{eq:bosonicmeasure}
\ee

%%%%%%%%%%%%%%%%%%%%%%%%%%%%%%
%%%%%%%%%%%%%%%%%%%%%%%%%%%%%%

\subsection{Fermion Zero Modes}
The fermion zero modes are goldstinos arising from the broken supersymmetries. When the monopole is in a single complex plane, the supersymmetry variations \C{susy} give
\bea
\delta \Psib^{P'} = (\Dslash_\mu \Zb_1)\etab^{1P'}
\eea
for $P'=2,3,4$, and by antisymmetry $\etab^{11}=0$. This implies there are $6$ broken supersymmetries, which after lowering the indices, are $\eta_{34},\eta_{23}, -\eta_{24}$, and the corresponding goldstinos are $\Psib^{2,3,4}$. This is in agreement with the fermionic equations of motion, [c.f. \C{fermions}], $\overline{\cDslash} \delta \Psib^{\,P'} = 0$ having zero modes for $P'=2,3,4$.
However, from \C{fermions} we see that $\Psi_1$ also has two zero modes as it obeys $\overline{\cDslash} \delta \Psi_{1}  = 0$, which does not have an interpretation as a Goldstino.

As pointed out in \cite{Hosomichi:2008ip}, things become clearer if we examine a generic background. Start a generic form of the scalar vev:
\be
Z^P = \begin{pmatrix} z^P  &0 \\ 0 & w^P \end{pmatrix} \quad \Zb_P = \begin{pmatrix} \ol{z}_P  &0 \\ 0 & \ol{w}_P \end{pmatrix}
\label{vevsd1}
\ee
We can use the $\SU(4)_R$-symmetry to rotate the vevs into a pair of complex planes. Denote the location of the M2-branes by complex 4-vectors: $\vec{z}=(z^1, \ldots, z^4)$ and $\vec{w} = (w^1, \ldots, w^4)$. Use the $SU(4)_R$-symmetry to set $z^{3,4}=w^{3,4}=0$. This leaves us with four complex or eight real parameters. Four of these may be eliminated by the remaining $\SU(2)_R\subset \SU(4)_R$ R-symmetry. A convenient (over)-parametrization is given by $\vec{z}=(z,d w^*,0,0)$ and $\vec{w}=(w,dz^*,0,0)$, which in terms of matrices is
\bea
Z^1 = \begin{pmatrix} z  &0 \\ 0 & w \end{pmatrix} \quad Z^2 = \begin{pmatrix} dw^*  &0 \\ 0 & dz^* \end{pmatrix}\label{vevs3}.
\label{vevsd2}
\eea
Demanding we preserve the diagonal matrix structure breaks the $\SU(2)_L\times\SU(2)_R$ gauge symmetry down to $\U(1)_D^2 \times \U(1)_b^2$. This vev leaves the $\U(1)_D^2$ unbroken but spontaneously breaks $\U(1)_b^2$  symmetry.  Let us write $\U(1)_b^2 = \U(1)(\sigma_3) \times \U(1)(\id)$. The former rotates the M2-branes by a relative phase; the latter by an overall phase. The angular momentum, corresponding to $\U(1)(\sigma_3)$ rotations, sources the monopole via the equations of motion, while the overall rotation $\U(1)(\id)$ decouples. To ensure the right transformation properties, we assign $d$ a charge $(0,+2)$ under $\U(1)(\sigma_3)\times\U(1)(\id)$, while $z$ has charge $(+1,+1)$ and $w$ charges $(-1,+1)$. Computing the $\SU(4)_R$ invariants
\bea
|\vec{z}|^2 &=& |z|^2 + |d|^2 |w|^2,\cr
|\vec{w}|^2 &=& |w|^2 + |d|^2 |z|^2,\cr
\vec{z} \cdot \vec{w}^* &=& z w^* + |d|^2 z^* w
\eea
we see that we may as well take $d$ to be real. This amounts to choosing a centre of mass for the relative angular separation of the branes -- moving the branes along by a constant overall phase doesn't change physics.  Furthermore, the phase of either $z$ or $w$ may be eliminated by an R-symmetry rotation and amounts to a rotation of the overall centre of mass of the monopole. This brings us down to four real degrees of freedom. With this choice of parametrization the background is given by a generalization of \C{monopole_1}:
\bea
Z^1 &=& aI_2, \quad \Zb_1 = (b\Phi(x) + cI_2),\cr
Z^2 &=& d \left(-b\Phi(x) +  cI_2\right), \quad \Zb_2 = ad^* I_2,\label{monopole2}
\eea
where recall from \eqref{monopolecoefficients}, \eqref{dconstant}, the coefficients $a,b,c,d$ are
\beqnn
a = \sqrt{zw}, \quad b=(|z|^2-|w|^2)/{\sqrt{zw}}, \quad c=(|z|^2+|w|^2)/{\sqrt{zw}}, \quad d = \frac{|z_2|^2 + |w_2|^2}{|z_1|^2 + |w_1|^2}.
\eeqnn
On the moduli space, the relevant fermionic supersymmetry variations are
\bea
\delta \Psib^{P'} = (\Dslash_\mu \Zb_1)\etab^{1P'} \quad \delta \Psi_{P'} = (\Dslash_\mu Z^2)\eta_{2P'}
\eea
with all other variations vanishing. There is an additional broken supersymmetry from $\Dslash Z^2 \ne 0$, so that the broken supersymmetries are $\{ \eta_{12}, \eta_{23}, \eta_{24}, \eta_{34}\}$. The Goldstinos arising from varying the background are
\bea
\delta \psi_1 &=& -(\Dslash Z^2) \eta_{12}, \cr
\delta \psib^2 &=& (\Dslash \Zb_1) \eta_{34}, \cr
\delta \psib^3 &=& -(\Dslash \Zb_1) \eta_{24}, \quad \delta \psi^4 = (\Dslash Z^2) \eta_{24} \cr
\delta \psib^4 &=& (\Dslash \Zb_1) \eta_{23}, \quad \delta \psi^3 = (\Dslash Z^2) \eta_{23}\label{fermionzero0}
\eea
We see the Goldstinos $\psib^3,\psi^4$ and $\psib^4,\psi^3$ are paired together, coming from the same supersymmetry variation. Its easy to see the linear combination $ d \psib^3 -  \psi_4$ and $d \psib^4 + \psi_3$ are preserved by the SUSY variations. Then, using \C{monopole2} a choice of polarisation for the goldstinos/zero modes is:
\be
\Psi^{(1)} = \psi_1, \quad \Psi^{(2)} = \psib^{\:2}, \quad \Psi^{(3)} =  Z^1  \psib^{\:3} + \Zb_2\psi_4, \quad \Psi^{(4)} =  Z^1\psib^{\:4} - \Zb_2\psi_3.\label{fermionzero1}
\ee
We will need the asymptotic form of these zero modes. Using \C{fermionzero0} and $D_\mu \Phi(x) \sim x_\mu / m x^3$ as $x/m \rightarrow \infty$ we find
\be
\begin{split}
\Psi^{(1)} ~=~& -4\pi b d \GLD \eta_{12}m^{-1}\sigma_3, \cr
\Psi^{(2)} ~=~& 4\pi b \GLD \eta_{34}m^{-1}\sigma_3,\cr
\Psi^{(3)} ~=~& -2k\GLD \eta_{24}\sigma_3,\cr
\Psi^{(4)} ~=~& 2k\GLD\eta_{23}\sigma_3.
\end{split}
\label{fermionzero2}
\ee
where $\GLD(x) = \frac{x_\mu\gamma^\mu}{ 4\pi x^3}$ is the free fermionic propagator, and we are suppressing spinor indices.

As for the bosonic zero-modes the fermionic zero-mode measure is fixed by normalizing the non-zero modes and letting the normalization constant descend. That is, we require
\be
N_F \int \prod_{P=1}^4 \mathcal{D} \psi_P \mathcal{D} \bar\psi^P \exp\left(-m\int d^3 x \bar\psi^P \psi_P \right) = 1,
\ee
to fix $N_F$. The zero-mode measure is then
\bea
d\mu_F &=& \left( mV\pi\right)^{-4}  \prod_{n=1}^4 (d\Psi^{(n)})^2 ,
\eea
where $\Psi^{(n)}$ are listed in \C{fermionzero1}. The measure should be $\SU(4)$ and gauge invariant up to breaking by the choice of vacuum. The eight fermion operator is constructed from a product of two four-fermion operators of which the possibilities are listed in \C{su4fermions}. When the monopole is in a single complex plane, the measure descends from the first operator. The generic case involves a linear combination of the two.

We now compute the fermionic Jacobian that arises when we convert to the fermionic collective coordinates $\eta_{PQ}$. For each zero mode listed in \C{fermionzero1}, there is a Jacobian
\be
J^{(i)}_\eta = \frac{1}{V} \int d^3x d^2 \eta \tr  \delta\Psi^{(i)}  \delta\Psi^{\,(i)},
\ee
so that the fermionic measure is given by
\be
 d\mu_F = \left( mV\pi\right)^{-4} \prod_{i=1}^4 (J_\eta^{(i)})^{-1}d^8 \eta
\ee
Computing the first Jacobian we find
\bea
J^{(1)}_\eta &=&\frac{1}{V} \int d^3x d^2 \eta_{12} \,\tr (\Dslash Z^2)^2 \eta_{12}^2\cr
&=& \frac{4\pi p}{mV} (bd)^2,
\eea
where $p$ is the monopole-instanton charge, the monopole mass $m$ is suitably extended for the vevs \C{monopole2} viz.
\be
m = \frac{2\pi}{k} a b(1+|d|^2).\label{monopolemass3}
\ee
and  $a,b,d$ are defined in \C{monopolecoefficients} and \eqref{dconstant}. Repeating this exercise for the four remaining zero modes we find
\bea
J^{(2)}_\eta &=& \frac{4\pi p}{mV} b^2,\cr
J^{(3)}_\eta &=& \frac{8\pi p}{mV} a^2b^2(1+|d|^2)^2,\cr
J^{(4)}_\eta &=& \frac{8\pi p}{mV}a^2b^2(1+|d|^2)^2,\label{fermionjacobian1}
\eea
Note that $b^2 d$ is invariant under the $\U(1)(\sigma_3)\times\U(1)(1)$ implying $(J_1 J_2)$, $J_3$, and $J_4$ are invariant under $\U(1)(\sigma_3)\times\U(1)(1)$, and as such the fermionic Jacobian $\prod_{i=1}^4 J_{(i)}$ is also invariant. This gives a zero mode measure of the form
\be
\int d\mu_F = \frac{1}{4 (2\pi pk)^4 (b^2 d)^2 m^2}\int d^8 \eta.
\label{fermionmeasure}
\ee

%%%%%%%%%%%%%%%%%%%%%%%%%%%%%%
%%%%%%%%%%%%%%%%%%%%%%%%%%%%%%

\newpage
\bibliographystyle{utphys}
\providecommand{\href}[2]{#2}\begingroup\raggedright\endgroup
\end{document}